\documentclass[11pt,preprint2]{aastex}
\def\obc {$\rm [OBC97] $}

\def\Msol {\hbox{M$_{\odot}$}}

\def\kms {\hbox{${\rm km\, s}^{-1}$}}
\def\acs {\hbox{$^{\prime\prime}$}}
\def\acm {\hbox{$^{\prime}$}}
\def\IRAM {IRAM 30m}
\slugcomment{Accepted by The Astrophysical Journal}

\shortauthors{O'Neil et al.}
\shorttitle{Further Discoveries of $^{12}$CO in LSBGs}

\begin{document}
\title{Further Discoveries of $^{12}$CO in Low Surface Brightness Galaxies}
\author{K. O'Neil}
\affil{NAIC/Arecibo Observatory, HC3 Box 53995, Arecibo, PR 00612 USA}
\email{koneil@naic.edu}

\author{E. Schinnerer\footnote{Work done while at California Institute of Technology}}
\affil{Jansky Fellow; NRAO, P.O. Box 0, Socorro, NM 87801 USA}
\email{eschinne@nrao.edu}

\and 

\author{P. Hofner}
\affil{Physics Department, U. of Puerto Rico at Rio Piedras, P.O. Box 23343, San Juan, PR 00931 USA}
\affil{NAIC/Arecibo Observatory, HC3 Box 53995, Arecibo, PR 00612 USA}
\email{hofner@naic.edu}

\begin{abstract}
Using the IRAM 30m telescope we have obtained seven new,
deep CO J(1$-$0) and J(2$-$1) observations of low surface brightness (LSB) galaxies.
Five of the galaxies have no CO detected to extremely low limits (0.1 -- 0.4 K km s$^{-1}$ at J(1$-$0)),
while two of the galaxies, UGC 01922 and UGC 12289, have
clear detections in both line transitions.  When these observations
are combined with all previous CO observations taken of LSB systems,
we compile a total of 34 observations, in which only 3 galaxies have
had detections of their molecular gas.  Comparing 
the LSB galaxies with and without CO detections to
a sample of high surface brightness (HSB) galaxies with CO observations
indicates that it is primarily the low density of baryonic matter within LSB galaxies
which is causing their low CO fluxes.  Finally, we note that one of the
massive LSB galaxies studied in this project, UGC 06968 (a Malin-1 `cousin'), has
upper limits placed on both M$_{H_2}$ and M$_{H_2}$/M$_{HI}$
which are 10--20 times lower than the {\it lowest} values found
for any galaxy (LSB or HSB) with similar global properties.
This may be due to an extremely low temperature and metallicity within UGC 06968,
or simply due to the CO distribution within the galaxy being too diffuse to be detected by the
IRAM beam.
\end{abstract}

\section{Introduction}
Despite more than a decade of study, the star formation processes within
low surface brightness (LSB) galaxies remain enigmatic. 
The general properties of LSB galaxies -- blue colors, high gas
mass-to-luminosity ratios, and low metallicities -- lead to the conclusion
that LSB systems are under-evolved compared to their high surface brightness
(HSB) counterparts.   When combined with the low gas density
(typically $\rm \rho_{HI}\;\leq\;10^{21}\;cm^{-2}$) and low
baryonic-to-dark matter content typical of LSB systems, the question 
can be raised not of why LSB galaxies are under-evolved, but instead
of how LSB systems ever form stars at all \citep{burkholder01,oneil00,bergvall99,
deblok98,roennback95, mcgaugh94b,vdhulst93,davies90}.

One of the primary methods for studying the star formation rate and
efficiency in galaxies is through study of the galaxies' ISM.
One mechanism for studying a galaxy's ISM is
through observing its CO content.  Numerous attempts have
been made looking for CO in LSB systems without success \citep{braine00,deblok98,schom90}.
If CO is as good a tracer of H$_2$ in LSB galaxies as it is in many HSB systems,
the failures in detecting CO would imply a lack of molecular gas within
LSB galaxies.  Alternatively, the previous non-detections could be a result
of a CO-to-H$_2$ ratio which is different in LSB systems than in, e.g. the Milky Way.
However, without any CO detections within LSB systems discerning between the 
two possibilities is difficult, if not impossible.

Recently our group made the first
CO detection in an LSB galaxy (\obc~P06-1) \citep{oneil00b}.  Significant though
this detection is, knowing the molecular content of one galaxy
is not enough to permit an understanding of the star formation
within LSB systems as a whole.  
Emboldened by the success of our earlier survey, and wishing to obtain a 
larger sample of CO measurements within LSB galaxies, we have observed another
set of seven LSB galaxies with the IRAM 30m telescope to look for the 
CO J($1-0$) and J($2-1$) lines.  The results of our survey
are presented here.  In order to understand better the molecular
properties of LSB galaxies as a whole, we also include the results
of all previously published CO observations of LSB systems and compare the molecular
properties of LSB galaxies with those of similar HSB systems.

A Hubble constant of H$_0$=75 km s$^{-1}$ Mpc$^{-1}$ is assumed throughout
this paper.

\begin{deluxetable}{lcccccccc}
\scriptsize
\tablecolumns{8}
\footnotesize
\tablewidth{0pt}
\tablecaption{Sources Observed in CO \label{tab:galobs}}
\tablehead{
\colhead{Name} & \colhead{RA}& \colhead{DEC}& \colhead{Int. Time}&
\colhead{$\sigma_{rms}^{1-0}$}& \colhead{$\sigma_{rms}^{2-1}$}& 
\colhead{Res.$\dagger$} &\colhead{Set-up$\ddagger$}\\
& \colhead{[J2000]}& \colhead{[J2000]}& \colhead{[min]}& \colhead{[mK]}&
\colhead{[\kms]}& \colhead{[mK]}
}
\startdata
UGC 01922	& 02:27:46.01 & 28:12:30.3& 730& 0.7& 0.8& 33& B\\
\obc\ C04-1	& 08:24:33.19 & 21:27:07.8& 310& 1.1& 2.5& 33& B\\
\obc\ N09-2	& 10:20:25.01 & 28:13:40.9& 440& 1.2& 2.0& 26& A\\
UGC 06968	& 11:58:44.63 & 28:17:23.2& 690& 0.5& 1.5& 33& B\\
UGC 12289	& 22:59:41.59 & 24:04:28.5& 360& 0.9& 1.0& 33& B\\
\obc\ P05-6	& 23:21:46.97 & 09:02:28.5& 640& 0.7& 1.3& 26& A\\
\obc\ P07-1	& 23:22:58.87 & 07:40:26.7& 480& 0.9& 1.2& 26& A\\
\enddata
\tablecomments{$\dagger$Smoothed resolution used for data analysis.\\
$\ddagger$See Section \ref{sec:CO_obs}\ for a description of the backend set-ups.}
\end{deluxetable}

\section{Observations}

\subsection{CO \label{sec:CO_obs}}

\begin{figure}[h]
\plottwo{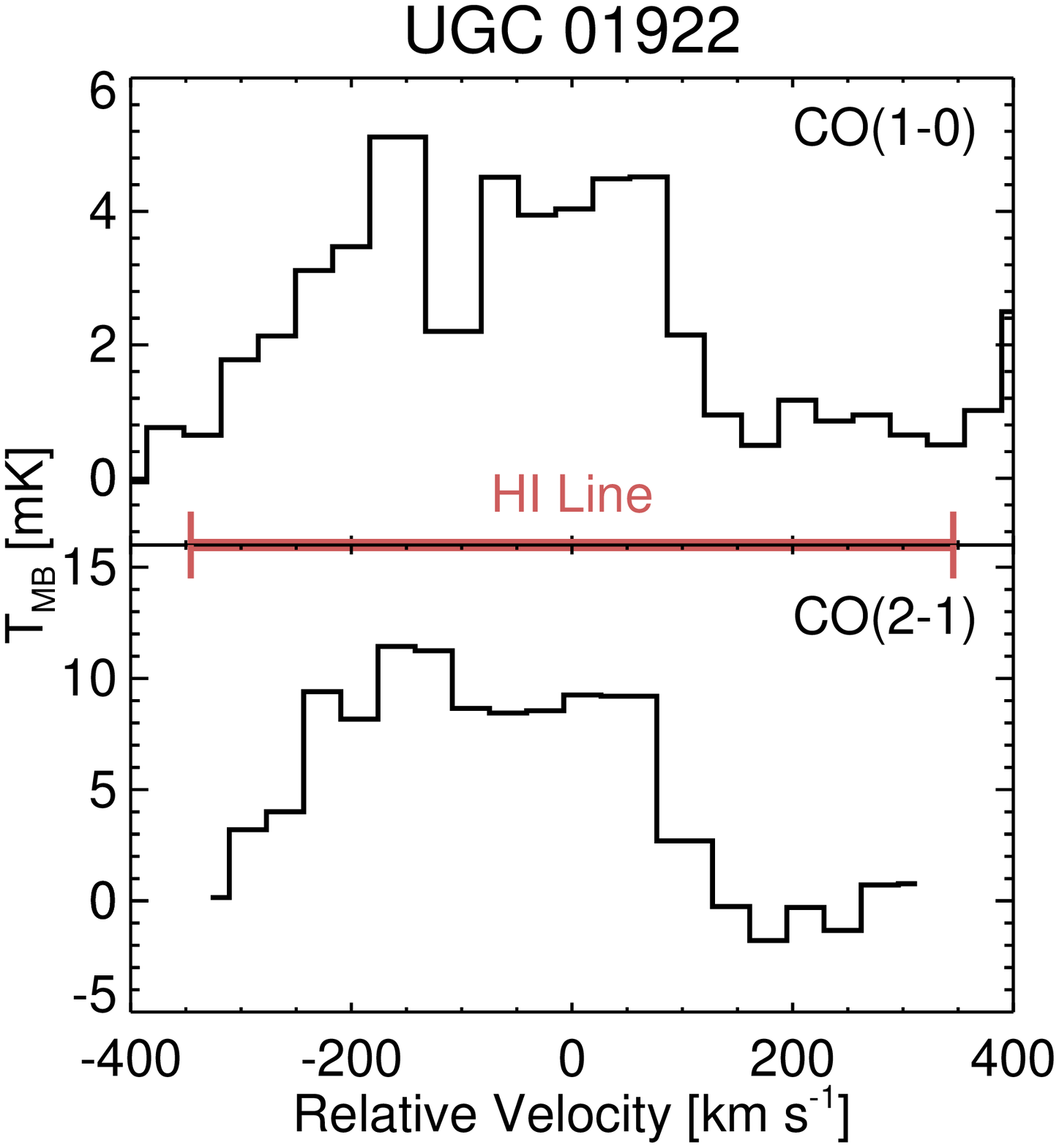}{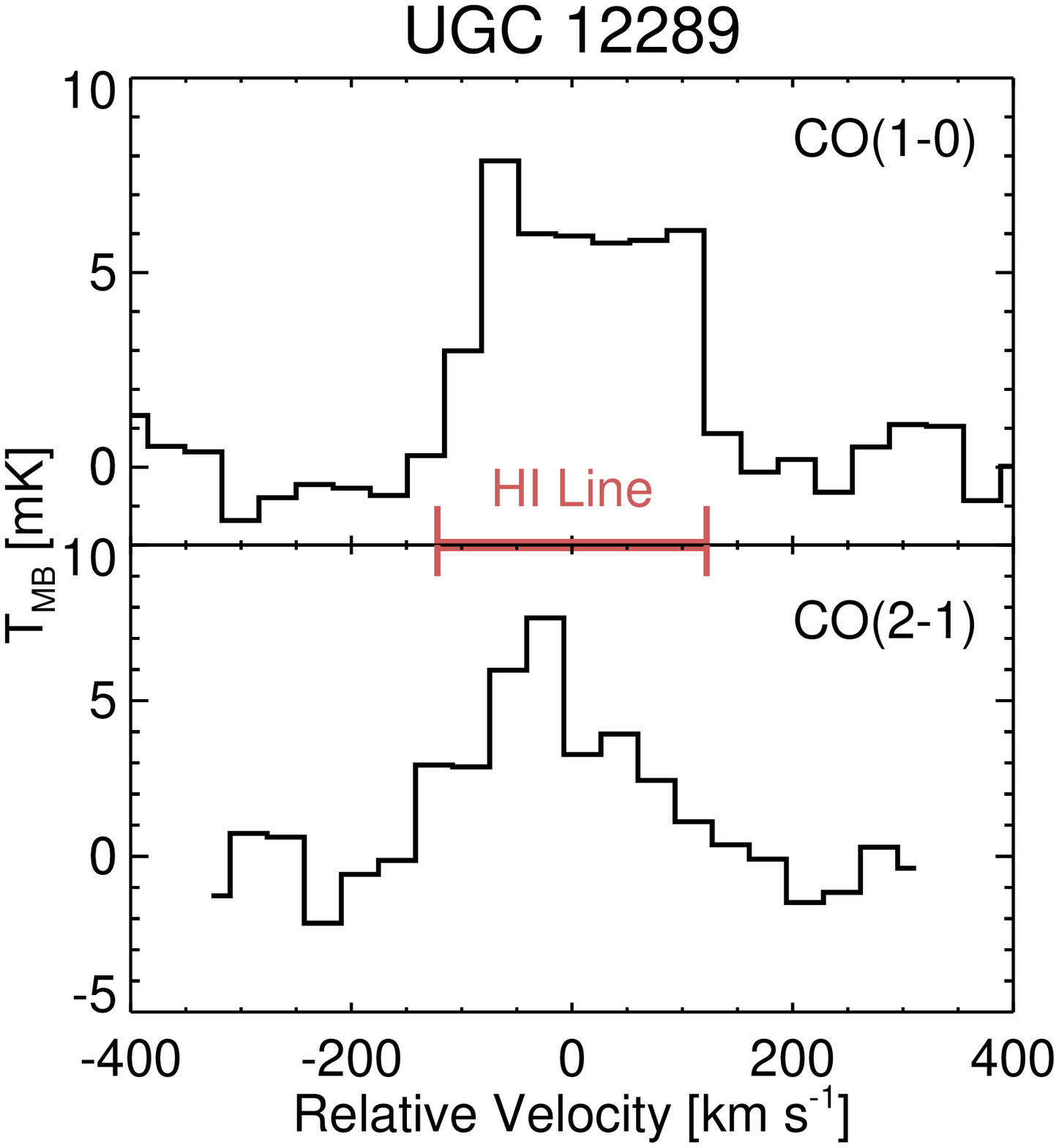}
\caption{IRAM 30m CO J(1$-$0) and J(2$-$1) spectra for UGC~01922 (left) and
UGC~12289 (right), the two galaxies in this survey with detectable amounts
of molecular gas.  The data have been smoothed to a resolution of 34 \kms.
The horizontal bars indicate the extent of the observed (uncorrected) \ion{H}{1}
velocity widths (at 20\% the peak HI intensity). \label{fig:codets}}
\end{figure}

\begin{figure}[h]
\plottwo{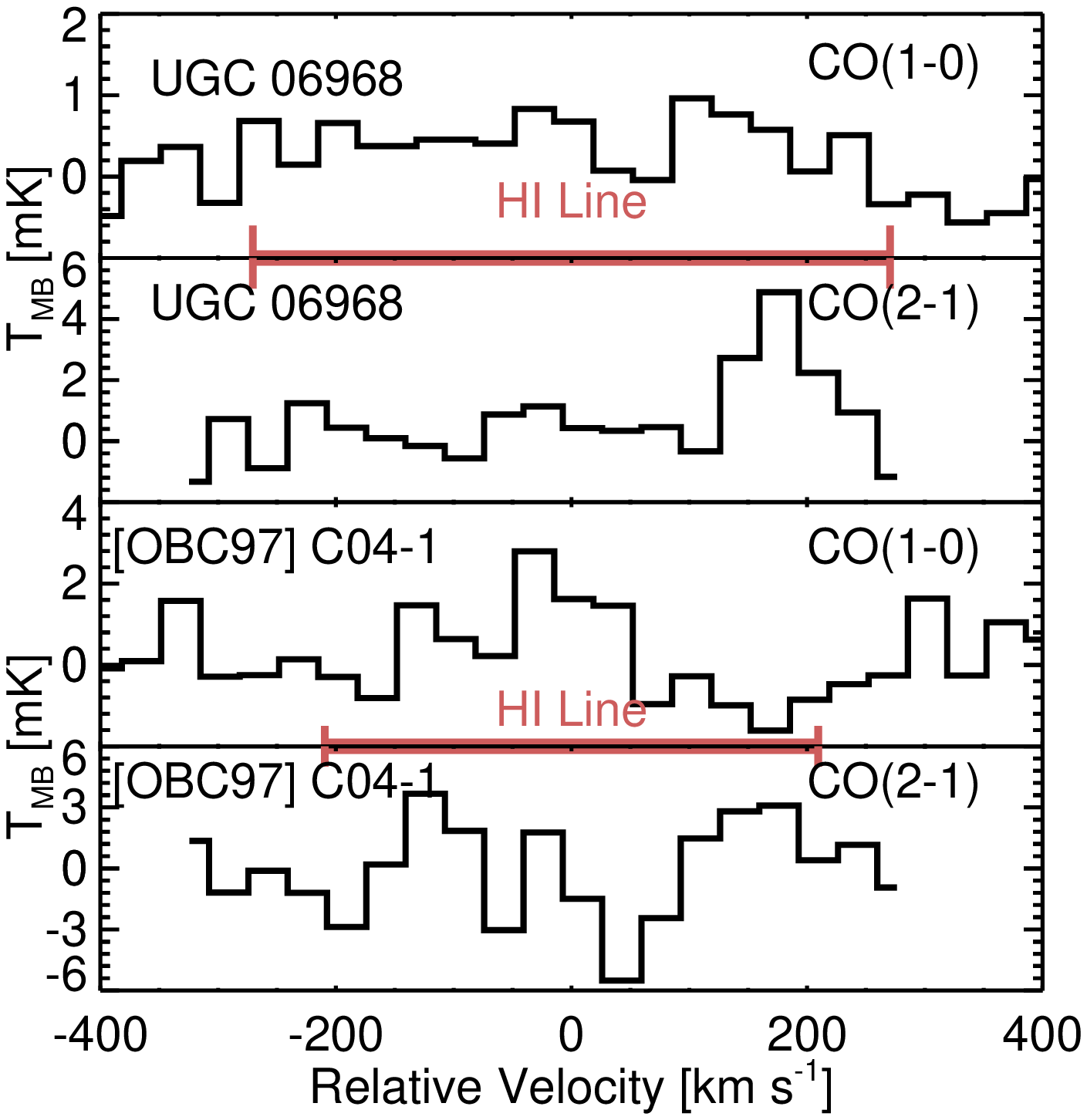}{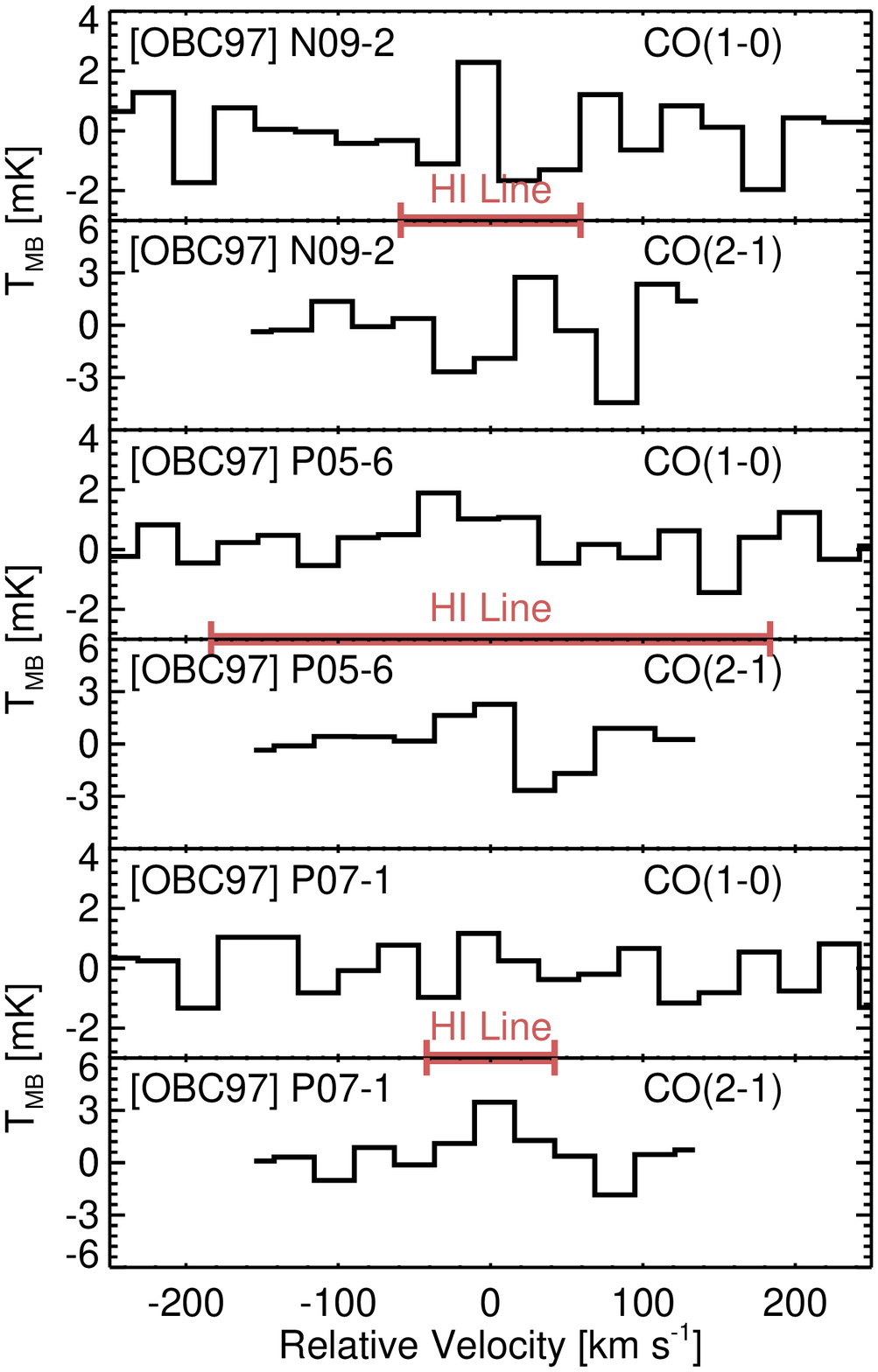}
\caption{IRAM 30m CO J(1$-$0) and J(2$-$1) spectra for the five 
galaxies in our survey for which we derived only upper limits on the 
CO line strengths.  The data have been smoothed to a resolution of 34 \kms.
The horizontal bars indicate the extent of the observed (uncorrected) \ion{H}{1}
velocity widths (at 20\% the peak HI intensity). \label{fig:conondets}}
\end{figure}

Based off previous observations of CO in LSB galaxies,
we chose to look for evidence of molecular gas
in LSB systems similar to the global properties of \obc~P06-1, the one LSB
galaxy with a CO detection \citep{oneil00b}.  Comparing the properties of
P06-1 with the other LSB galaxies observed at CO, it is clear that P06-1
is one of the more massive systems studied (log(M$_{HI}$/M$_\odot$) = 10,
W$_{20}$ = 460 \kms -- Table~\ref{tab:props}), and also one whose optical
spectra indicates some nuclear activity (Section 3.2).  With this in mind,
we chose to focus the observations on massive LSB systems, some with and some
without AGN/LINER characteristics.  The galaxies for this sample were chosen from three separate
sources -- the massive LSB galaxy catalog of \citet{oneil02}, the
AGN LSB list provided in \citet{schom98} of LSB galaxies with
AGN/LINER cores, and the LSB galaxy catalog of \citet{oneil00}.  A full list of the galaxies
observed and their properties are given in Tables~\ref{tab:galobs} and \ref{tab:props}, and
the results of the observations are given in Table~\ref{tab:COprops}.
The resultant spectra are shown in Figures~\ref{fig:codets} and \ref{fig:conondets}.

The CO J(1--0) and J(2--1) rotational transitions were observed with
the \IRAM\ telescope in the period from 25-28 May, 2001.
Table~\ref{tab:galobs} lists the adopted positions (determined using the
digitized Palomar sky survey plates and accurate to 2-3\acs) and heliocentric velocities (adopted from
the published values obtained from 21-cm HI observations) for our target sources.
The beams were centered on the nucleus of each galaxy.
Pointing  and focus were checked every hour
and pointing was found to be better than $5^{\prime\prime}$.
For each source both transitions were observed
simultaneously with two receivers.
Two different setups were used for the filterbanks, depending upon the
known HI line widths of the individual galaxy.  For those
galaxies known to have HI line widths less than 300 \kms\
both receivers used a 256 channel filterbank with channel width 
of 3 MHz (667 \kms\ bandwidth, 2.6 \kms\ resolution at 3mm) -- setup A.
For the sources which are known to have wider HI lines the
1mm lines used a 512 channel filter bank with 1 MHz channels for each polarizations,
while the 3mm line used the auto-correlator with 512 MHz bandwidth
and 1MHz channels (1331 \kms\ bandwidth, 2.6 \kms\ resolution at 3mm) -- setup B.
Table~\ref{tab:galobs} gives the set-up
used for the individual galaxies.
The FWHM of the IRAM beam is 22\acs\ and 11\acs\ for the 3mm and 1mm beams, respectively.

All observations used the wobbling secondary with
the maximal beam throw of $240^{\prime\prime}$.
The image side band rejection ratios were measured
to be $>30$dB for the $3\,$mm SIS receivers and
$>12$dB for the $1.3\,$mm SIS receivers.  The data were
calibrated using the standard chopper wheel technique \citep{kut81}
and are reported in main beam brightness temperature T$_{MB}$.
Typical system temperatures during the observations
were 170--190K and 350--450K in the $3\,$mm and $1.3\,$mm band, respectively.
All data reduction was done using CLASS -- the Continuum and Line Analysis
Single-dish Software developed by the Observatoire de Grenoble and IRAM \citep{buisson02}.

\subsection{Optical Spectra}

In order to further explore the properties of the studied galaxies' nuclear regions
(where the IRAM beam was centered), optical spectra of the nuclei of
a sample of LSB galaxies which have been
observed in CO were taken with the Palomar Observatory 5m
telescope{\footnote{Observations at the Palomar Observatory were made as a part
of the cooperative agreement between Cornell University and the California Institute of
Technology}} in October 2000 (the \obc~galaxies) and November 2001
(the UGC galaxies).  The observations
used the Double Spectrograph with both the blue and red
cameras.  The long slit (2\acm) was used and set to a 2" aperture
during all observations. Both cameras are 1024$\times$1024 TEK CCD cameras, giving
3.37\AA\ pixel$^{-1}$/0.624\acs\ pixel$^{-1}$ and 2.48\AA\ pixel$^{-1}$/0.48\acs\ pixel$^{-1}$
for the blue and red cameras, respectively.  As the blue CCD has
moderate focusing problems across the full wavelength range, the blue focus was
optimized for [\ion{O}{3}]$\lambda$3727.

During the October 2000 
run a 5500\AA\ dichroic filter was used to split the light to the blue
and red CCD cameras.  A 316 lines mm$^{-1}$ diffraction grating (blazed at 7150\AA) 
was used  for the red camera and the 300 lines mm$^{-1}$
diffraction grating (blazed at 3990\AA) was used for the blue camera during this run.
In November 2001 the 5200 \AA\ dichroic filter was used with
the 600 lines mm$^{-1}$ diffraction grating (blazed at 3780\AA) on the blue camera
and the 316 lines mm$^{-1}$ diffraction grating (blazed at 7150\AA) on the red camera.
The seeing averaged between 1.5 -- 2.1\acs\ for the observations.

\begin{deluxetable}{lcccccccccc}
\tabletypesize{\scriptsize}          
\tablecolumns{10}
\footnotesize          
\tablewidth{0pt}          
\tablecaption{Known Properties of All LSB Galaxies Observed at CO
\label{tab:props}}
\tablehead{
\colhead{Name}  & \colhead{Type} & \colhead{$\mu_B(0)$} & \colhead{B$^a_T$} & \colhead{D$^a$}&
\colhead{B$-$V} &    
\colhead{\it  i} & \colhead{log(${L_{FIR}}\over{L_\odot}$)$^b$} & \colhead{log(${M_{HI}}\over{M_\odot}$)} &
\colhead{$\rm v_{HEL}^{HI}$} & \colhead{$\rm w_{20_{cor}}^{HI}$$^c$} \\      
\colhead{} & \colhead{}  & \colhead{[mag arcsec$^{-2}$]}& \colhead{[mag]}& \colhead{[kpc]} & \colhead{}&    
\colhead{[$^\circ$]} & \colhead{}  & \colhead{}  & \colhead{[km s$^{-1}$]}  & \colhead{[km s$^{-1}$]}
}          
\startdata          
\cutinhead{Detections  -- This Paper}      
UGC 01922    &S?$^1$ &\nodata$^d$&-19.8& 59& \nodata  &38$^2$  &10.6 &10.33$^3$& 10894$^3$ & 1120$^3$ \\
UGC 12289    &Sd$^1$ &23.3$^4$ &-19.7&   57& \nodata   &22$^2$  &10.7 &10.13$^3$& 10160$^3$ & 488$^3$\\
\cutinhead{Previous  Detection}        
\obc\ P06-1  &Sd$^5$ &23.2$^6$     &-18.6& 29& 0.9$^7$ &70$^6$  &$<$10.2 &9.87$^5$&  10882$^5$ & 458$^5$ \\
\cutinhead{Non-Detections  -- This Paper}      
\obc\ P07-1  &Sc$^5$ &23.3$^6$     &-16.3& 16&\nodata &75$^6$  &$<$9.4  &8.21$^5$&  3471$^5$  & 87$^5$ \\
\obc\ N09-2  &Im$^5$ &23.5$^6$     &-16.2& 13& 0.8$^7$ &47$^6$  &$<$9.7  &10.03$^5$& 7746$^5$  & 161$^5$ \\
\obc\ P05-6  &Sbc$^5$ &23.4$^6$    &-16.5& 17& 0.7$^7$ &75$^6$  &$<$9.3  &9.24$^5$&  3667$^5$  & 380$^5$ \\
\obc\ C04-1  &Im$^5$ &23.4$^6$	   &-17.3& 14& 0.5$^7$ &39$^6$	&$<$9.9  &9.74$^5$&  7905$^5$  & 419$^5$ \\
UGC 06968      &Sc$^1$ &\nodata$^d$&-21.1& 48& 1.0$^{8}$ &71$^2$ &$<$9.8 &10.30$^{9}$& 8232$^{9}$ & 574$^{9}$ \\
\cutinhead{Previous  Non-Detections}         
LSBC D575-05  &dI$^{10}$ &(22.5)$^e$  &-12.2&2.0&\nodata  &60$^{11}$  &$<$7.2  &7.47$^{12}$& 419$^{12}$ & 27$^{12}$ \\ 
LSBC F571-V2  &Im$^{13}$ &\nodata     &\nodata&3.3&\nodata  &37$^{10}$  &$<$7.8  &8.06$^{14}$& 955$^{14}$ & 53$^{14}$ \\
LSBC F638-3   &Sm$^{15}$ &\nodata     &-16.4&6.1&\nodata  &46$^{11}$  &$<$8.9  &8.88$^{15}$& 3160$^{15}$ & 58$^{15}$ \\
LSBC F583-5   &Sm$^{10}$ &23.5$^{16}$ &-14.7& 24&0.6$^{16}$ &65$^{16}$ &$<$8.9  &9.11$^{17}$& 3261$^{17}$ & 28$^{17}$ \\
LSBC F651-2   &dI$^{15}$ &\nodata     &-15.5&4.9&\nodata &46$^{11}$  &$<$8.3  &8.73$^{15}$& 1789$^{15}$ & 70$^{15}$ \\
LSBC F721-1   &Sdm$^{15}$ &\nodata    &-18.9& 17&\nodata &41$^{11}$  &$<$9.7  &9.52$^{15}$& 7210$^{15}$ & 98$^{15}$ \\
LSBC F571-V1  &Im$^{15}$ &24.0$^{18}$ &-15.9& 20&0.4$^{18}$ &35$^{18}$ &$<$9.5  &9.07$^{15}$& 5719$^{15}$ & 113$^{15}$ \\
\obc\ P05-5 &Im$^5$    &24.4$^6$    &-14.5&  5.7& 1.2$^7$    &49$^6$    &$<$9.1  &7.78$^5$&    3177$^6$    & 94$^6$ \\
LSBC F636-1   &Sm(r)$^{15}$ &\nodata  &-17.5& 13&\nodata  &52$^{11}$  &$<$9.3  &9.27$^{15}$& 4302$^{15}$ & 95$^{15}$ \\
LSBC F562-V2  &S:$^{10}$ &\nodata     &-18.1& 12&\nodata  &20$^{10}$  &$<$9.4  &9.59$^{15}$& 6330$^{15}$ & 156$^{15}$\\
LSBC F563-V1  &Im$^{15}$ &23.4$^{18}$ &-15.8& 6.1&0.6$^{18}$ &60$^{18}$ &$<$9.3  &8.73$^{15}$& 3931$^{15}$ & 90$^{15}$ \\
LSBC F561-1   &Sm$^{15}$ &23.2$^{18}$ &-17.5& 19&0.5$^{18}$ &24$^{18}$ &$<$9.4  &8.81$^{15}$& 4810$^{15}$ & 160$^{15}$\\
LSBC F574-10  &Sm$^{10}$ &\nodata     &-12.8&3.0&\nodata  &72$^{10}$  &$<$7.8  &8.23$^{15}$& 863$^{15}$  & 86$^{15}$ \\
LSBC F585-V1  &dI$^{15}$ &23.4$^{16}$ &-13.7&3.1&\nodata  &41$^{15}$ &$<$8.5  &8.11$^{13}$& 1985$^{13}$ & 126$^{13}$ \\
\obc\ C04-2 &Im$^5$ &24.0$^6$         &-16.5& 13& 1.3$^7$    &67$^6$    &$<$9.5  &8.82$^5$&    5168$^5$    & 111$^5$ \\
LSBC F563-V2 &Irr$^{15}$ &22.0$^{16}$ &-17.3& 15&0.5$^{16}$ &40$^{16}$ &$<$9.2  &9.31$^{15}$& 4311$^{15}$ & 184$^{15}$ \\
LSBC F563-1   &Im$^{15}$ &23.6$^{18}$ &-16.6& 14&0.9$^{18}$ &25$^{18}$ &$<$9.0  &9.04$^{14}$& 3495$^{14}$ & 252$^{14}$ \\
LSBC F571-5   &Sm$^{15}$ &23.7$^{18}$ &-16.0& 17&0.3$^{18}$ &26$^{18}$ &$<$9.3  &9.24$^{15}$& 4253$^{15}$ & 260$^{15}$ \\
\obc\ C05-3 &Im$^5$    &23.9$^6$    &-17.3&   20& 1.3$^7$    &51$^6$    &$<$10.0 &9.41$^5$&    12940$^5$   & 167$^5$\\
LSBC F585-3   &Sm$^{15}$ &\nodata     &-16.5& 13&\nodata  &74$^{10}$ &$<$9.0  &9.20$^{13}$& 3100$^{13}$ & 154$^{13}$ \\
LSBC F568-V1  &Sc$^{15}$ &23.2$^{18}$ &-17.2& 16&0.6$^{18}$ &40$^{18}$ &$<$9.5  &9.39$^{15}$& 5768$^{15}$ & 249$^{15}$ \\
LSBC F583-1 &Sm/Irr$^{15}$ &24.1$^{19}$&-15.9$^{14}$&8.8&\nodata &63$^{19}$ &$<$8.7 &9.18$^{15}$& 2264$^{15}$ & 181$^{15}$ \\
LSBC F582-V1 &Ring$^{10}$ &\nodata    &-18.0&9.1&\nodata  &20$^{10}$ &$<$10.0 &9.85$^{15}$& 11695$^{15}$ & 356$^{15}$ \\
LSBC F564-V2 &Im$^{15}$ &\nodata     &\nodata&9.5&\nodata  &29$^{10}$ &$<$9.0  &8.91$^{13}$& 3060$^{13}$ & 358$^{13}$ \\
LSBC F582-2  &Sbc$^{15}$ &\nodata    &\nodata&41&\nodata  &66$^{10}$ &$<$9.6  &9.99$^{15}$& 7043$^{15}$ & 310$^{15}$ \\
Malin 1      &S$^{20}$  &26.4$^{20}$ &-21.4$^{20}$&240&0.9$^{20}$ &20$^{20}$ &$<$11.0 &10.6$^{20}$&  24733$^{20}$ & 710$^{20}$\\
\enddata           
\tablecomments{
$^a$Unless otherwise noted magnitudes and diameters are taken from NED, the NASA Extragalactic Database.
$^b$FIR luminosities are given by $\log(L_{FIR})\;=\;46.73\;+\;2log[z(1+z)]\;+\;log[2.58f_{60}+f_{100}]$
where $\rm f_{60}\;and\;f_{100}$ are in Jy and $\rm L_{FIR}$ is in erg s$^{-1}$.  Fluxes were
obtained from the Infrared Astronomical Satellite (IRAS) data.
$^c$Velocity widths are corrected for inclination using $w_{20_{corr}}^{HI}\;=\;w_{20}^{HI}/sin\;i$.
To avoid over-correcting the velocity widths, the smallest inclination value used for the purpose of correcting the
velocity widths is 30$^\circ$.
$^d$The classification of this galaxy as an LSB galaxy is from \citet{schom98}.
$^e$The reported central surface brightness is through the V (not B) band \citep{pildis97}.
}
\tablerefs{
$^1$~Schombert (1998);
$^2$~Nilson (1973); 
$^3$~Giovanelli \& Haynes (1985); 
$^4$~Data taken with the Pine Mountain Observatory telescope.
PMO is operated by the University of Oregon Physics Department with help from
the Friends of Pine Mountain Observatory.  Image obtained courtesy of G. Bothun.;
$^5$~O'Neil, Bothun, \& Schombert (2000);
$^6$~O'Neil, Bothun, \& Cornell (1997a);
$^7$~O'Neil, et.al (1997b);
$^{8}$~Boselli \& Gavazzi (1994);
$^{9}$~Gavazzi (1987);
$^{10}$~Schombert \& Bothun (1988);
$^{11}$~Garnier, et.al (1996);
$^{12}$~Eder \& Schombert (2000);
$^{13}$~Schombert, et.al (1990);
$^{14}$~de Blok, McGaugh, \& van der Hulst (1996);
$^{15}$~Schombert, et.al (1992);
$^{16}$~McGaugh \& Bothun (1994);
$^{17}$~Huchtmeier, Karachentsev, Karachentseva (2000);
$^{18}$~de Blok, van der Hulst \& Bothun (1995);
$^{19}$~de Blok \& McGaugh (1997);
$^{20}$~Impey \& Bothun (1989)
}
\end{deluxetable}          

\thispagestyle{empty}
\begin{deluxetable}{lcccccc}
\tabletypesize{\scriptsize}
\tablecolumns{7}
\footnotesize
\tablewidth{0pt}
\tablecaption{CO Properties of All LSB Galaxies Observed at CO
\label{tab:COprops}}
\tablehead{
\colhead{Galaxy} & \colhead{Line}&
\colhead{$\rm \int{T_{MB}dv}\dagger$}  &\colhead{$\rm v_{HEL}$}&
\colhead{Width} & \colhead{$\rm log(M_{H_2}/M_\odot$)$\ddagger$}& \colhead{$\rm M_{H_2}/M_{HI}$}\\
& & \colhead{[K km s$^{-1}$]}& \colhead{[km s$^{-1}$]} & \colhead{[km s$^{-1}$]} &&
}
\startdata
UGC 01922        & 1$-$0& 1.38& 10795& 404& 9.2& 0.07\\
UGC 01922        & 2$-$1& 2.96& 10802& 403& 8.9& 0.04\\
UGC 12289        & 1$-$0& 1.16& 10162& 200& 9.0& 0.07\\
UGC 12289        & 2$-$1& 0.69& 10185& 201& 8.2& 0.01\\
\cutinhead{Previous Detection}
\obc\ P06-1$^1$  & 1$-$0&0.95& 10904& 302& 8.8& 0.09\\
\obc\ P06-1$^1$  & 2$-$1&1.14& 10903& 216& 8.3& 0.03\\
\cutinhead{Non-detections - This Paper}
\obc\ P07-1      & 1$-$0&$<$0.13&\nodata&\nodata&  $<$7.0& $<$0.006\\
\obc\ P07-1      & 2$-$1&$<$0.032&\nodata&\nodata& $<$5.7& $<$0.003\\
\obc\ N09-2      & 1$-$0&$<$0.18&\nodata&\nodata& $<$7.8& $<$0.006\\
\obc\ N09-2      & 2$-$1&$<$0.33&\nodata&\nodata& $<$7.5& $<$0.003\\
\obc\ P05-6      & 1$-$0&$<$0.22&\nodata&\nodata& $<$7.2& $<$0.009\\
\obc\ P05-6      & 2$-$1&$<$0.39&\nodata&\nodata& $<$6.9& $<$0.005\\
\obc\ C04-1      & 1$-$0&$<$0.37& \nodata&\nodata& $<$8.1& $<$0.02\\
\obc\ C04-1      & 2$-$1&$<$0.88&\nodata&\nodata& $<$7.9& $<$0.01\\
UGC 06968        & 1$-$0&$<$0.21&\nodata&\nodata& $<$7.9& $<$0.004\\
UGC 06968        & 2$-$1&$<$0.58& \nodata&\nodata& $<$7.8& $<$0.003\\
\cutinhead{Previous Non-Detections}
LSBC F575-3$^2$  & 1$-$0&$<$0.14& \nodata&\nodata& $<$6.1& $<$0.05\\
LSBC F571-V2$^2$ & 1$-$0&$<$0.19& \nodata&\nodata& $<$7.0& $<$0.09\\
LSBC F638-3$^2$  & 1$-$0&$<$0.22& \nodata&\nodata& $<$8.1& $<$0.2\\
LSBC F583-5$^2$  & 1$-$0&$<$1.20& \nodata&\nodata&  $<$7.9& $<$0.07\\
LSBC F651-2$^2$  & 1$-$0&$<$0.22& \nodata&\nodata& $<$7.6& $<$0.08\\
LSBC F721-1$^2$  & 1$-$0&$<$0.19& \nodata&\nodata& $<$8.7& $<$0.2\\
LSBC F571-V1$^3$ & 2$-$1&$<$0.20& \nodata&\nodata&  $<$7.6& $<$0.2\\
\obc\ P05-5$^1$  & 1$-$0&$<$0.12& \nodata&\nodata& $<$7.6& $<$0.7\\
\obc\ P05-5$^1$  & 2$-$1&$<$0.21& \nodata&\nodata& $<$7.3& $<$0.3\\
LSBC F636-1$^2$  & 1$-$0&$<$0.30& \nodata&\nodata& $<$8.5& $<$0.2\\
LSBC F562-V2$^2$ & 1$-$0&$<$0.33& \nodata&\nodata& $<$8.9& $<$0.2\\
LSBC F563-V1$^3$ & 2$-$1&$<$0.28& \nodata&\nodata&  $<$7.4& $<$0.2\\
LSBC F561-1$^1$  & 1$-$0&$<$0.93& \nodata&\nodata&  $<$8.3& $<$0.2\\
LSBC F574-10$^2$ & 1$-$0&$<$0.46& \nodata&\nodata& $<$7.3& $<$0.1\\
LSBC F585-V1$^2$ & 1$-$0&$<$1.17& \nodata&\nodata&  $<$7.7& $<$0.2\\
\obc\ C04-2$^1$  & 1$-$0&$<$0.096&\nodata&\nodata& $<$7.2& $<$0.02\\
LSBC F563-V2$^2$ & 1$-$0&$<$0.42& \nodata&\nodata& $<$8.6& $<$0.2\\
LSBC F563-1 $^2$ & 1$-$0&$<$0.53& \nodata&\nodata&  $<$8.0& $<$0.05\\
LSBC F571-5$^2$  & 1$-$0&$<$0.35& \nodata&\nodata& $<$8.5& $<$0.2\\
\obc\ C05-3$^1$  & 1$-$0&$<$0.14& \nodata&\nodata& $<$8.1& $<$0.05\\
LSBC F585-3$^2$  & 1$-$0&$<$0.66& \nodata&\nodata& $<$8.1& $<$0.04\\
LSBC F568-V1$^3$ & 2$-$1&$<$0.19& \nodata&\nodata&  $<$7.7& $<$0.05\\
LSBC F583-1$^2$  & 1$-$0&$<$0.49& \nodata&\nodata&  $<$8.1& $<$0.08\\
LSBC F582-V1$^2$ & 1$-$0&$<$0.47& \nodata&\nodata& $<$9.5& $<$0.4\\
LSBC F564-V2$^1$ & 1$-$0&$<$0.70& \nodata&\nodata&  $<$8.2& $<$0.1\\
LSBC F582-2$^2$  & 1$-$0&$<$0.54& \nodata&\nodata& $<$9.2& $<$0.2\\
Malin 1$^4$      & 1$-$0&$<$0.15& \nodata&\nodata& $<$9.4& $<$0.06\\
Malin 1$^4$      & 2$-$1&$<$0.35& \nodata&\nodata& $<$8.7& $<$0.01\\
\enddata
\tablecomments{$\dagger$Non-detection limits are
$I_{CO}\:<\:3{T_{MB}}{v^{20}_{HI}\over{\sqrt{N}}}$ (N = the number of channels,
T$_{MB}$ is the 1$\sigma$ rms main beam temperature).\\
$\ddagger$As described in Section 3, conversion to M$_{H_2}$ was done using
$\rm N(H_2)/\int T(CO)dv=3.6\times 10^{20}\;cm^{-2}/(K\;km\;s^{-1})$}
\tablerefs{$^1$O'Neil, Hofner, \& Schinnerer (2000); $^2$Schombert, et.al (1990);
$^3$de Blok \& van der Hulst (1998); $^4$Braine, Herpin, \& Radford (2000)}
\end{deluxetable}

\ptlandscape{
\begin{deluxetable}{lcccccccccc}
\tabletypesize{\scriptsize}
\tablewidth{0pt}
\tablecaption{Observed Optical Lines\label{tab:optical_obs}}
\tablehead{
\colhead{Galaxy} &
\multicolumn{2}{c}{H$\beta$} &
\multicolumn{2}{c}{\ion{O}{2} [$\lambda$3727]} &
\multicolumn{2}{c}{\ion{Ca}{2}(K) [$\lambda$3934]}&
\multicolumn{2}{c}{H7+\ion{Ca}{2}(H) [$\lambda$3969]}&
\multicolumn{2}{c}{G band [$\lambda$4306]}\\
& \colhead{EW} & \colhead{flux} 
& \colhead{EW} & \colhead{flux} 
& \colhead{EW} & \colhead{flux}  
& \colhead{EW} & \colhead{flux}  
& \colhead{EW} & \colhead{flux}  
}
\startdata
\obc\ C04-2&     8.0&     2.9&      27&      12& \nodata& \nodata& \nodata& \nodata& \nodata& \nodata \\
\obc\ P05-5& \nodata& \nodata& \nodata& \nodata& \nodata& \nodata& \nodata& \nodata& \nodata& \nodata \\
\obc\ P05-6& \nodata& \nodata&     9.0&     9.1& \nodata& \nodata& \nodata& \nodata& \nodata& \nodata \\
\obc\ P06-1&    -4.0&    -3.4&     4.1&     3.7&    -8.2&    -6.1&    -12 &   -8.8 & \nodata& \nodata \\
UGC 01922  &     3.2&      19&     9.0&      16&    -8.8&    -13 &    -6.1&   -11  &    -4.2&   -25   \\
UGC 12289  & \nodata& \nodata& \nodata& \nodata&    -8.3&    -3.2&    -8.1&   -3.5 &    -5.3&   -3.4 \\
\enddata
\tablecomments{All errors are within 15\%.
EW is in \AA. Flux is in units of 10$^{-16}$ erg s$^{-1}$ cm$^{-2}$ \AA$^{-1}$.
Values reported as negative are in absorption.
}
\end{deluxetable}
}
\addtocounter{table}{-1}
\ptlandscape{
\begin{deluxetable}{lccccccccc}
\tabletypesize{\scriptsize}
\tablewidth{0pt}
\tablecaption{Observed Optical Lines, {\it cont}}
\tablehead{
\colhead{Galaxy} &
\multicolumn{2}{c}{\ion{Na}{1} [$\lambda\lambda$5889,5895]}&
\multicolumn{2}{c}{H$\alpha$ [$\lambda$6563]} &
\multicolumn{2}{c}{\ion{N}{2} [$\lambda\lambda$6548,6583]} &
\multicolumn{2}{c}{\ion{S}{2} [$\lambda\lambda$6717,6731]} &
\colhead{c}\\
& \colhead{EW} & \colhead{flux}  
& \colhead{EW} & \colhead{flux}  
& \colhead{EW} & \colhead{flux} 
& \colhead{EW} & \colhead{flux}
}
\startdata
\obc\ C04-2& \nodata& \nodata& 40.&  13.&     6.0&    2.1&  20.&7.0&     0.52\\
\obc\ P05-5& \nodata& \nodata& 3.7& 0.63& \nodata& \nodata& 1.8&0.30& \nodata\\
\obc\ P05-6& \nodata& \nodata& 11.& 8.5 & \nodata& \nodata& 2.9& 1.9& \nodata\\
\obc\ P06-1&    -2.5&   -3.6 & 3.1& 5.5 &     2.1&    3.8 & 2.7& 6.2& \nodata\\
UGC 01922  &    -4.8& -41  & 91.& 120 &      130 &    170 & 140& 190&     0.36\\
UGC 12289  &    -3.3& -5.4 & 4.5& 6.7 &       6.2&    11. & 2.0& 2.9& \nodata\\
\enddata
\tablecomments{All errors are within 15\%.
EW is in \AA. Flux is in units of 10$^{-16}$ erg s$^{-1}$ cm$^{-2}$ \AA$^{-1}$.
Values reported as negative are in absorption.
The reddening coefficient c=1.28E($B-V$).
{\it Note that if no reddening coefficient is given, the data is uncorrected for Extragalactic
extinction.}}
\end{deluxetable}
}

\clearpage
The galaxies were acquired through blind offsets from nearby stars.  Positions were obtained
through a combination of both the HST Guide Star Catalog and the second generation Palomar
Survey Data.   Based off the studies of \citet{vanz98}
offsets were kept to less than 20\acm\ and guide stars were chosen to have m$_B$$<$11 to insure
accurate offsets and guiding.  To reduce light losses due to atmospheric refraction, the slit
position angle was set close to the parallactic angle during observations. 
The slits were placed across the nucleus of each galaxy.  The integration
times were 1200s per frame.  The total (on source) observing times were 120, 260, 220, 300, 80, and 120
minutes for \obc~C04-2, \obc~P05-5, \obc~P05-6, \obc~P06-1, UGC~01922, and UGC~12289, 
respectively. 

\begin{figure}[h]
\plotone{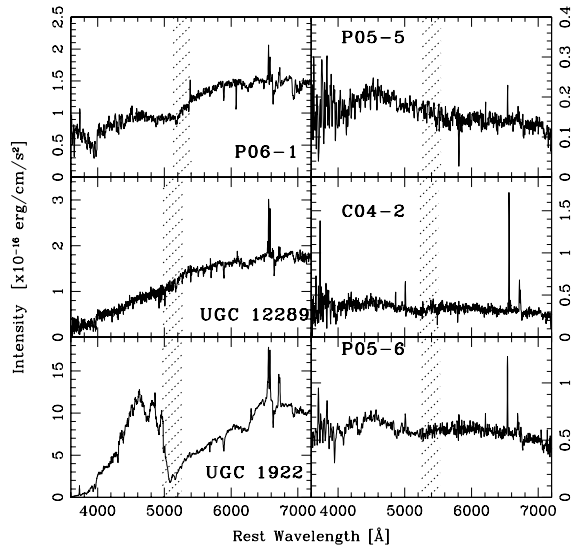}
\caption{Uncorrected optical spectra taken by the Palomar 5m telescope for
a sample of LSB galaxies with studied CO properties. The diagonal lines mark the boundary where
the dichroic filter transmission falls by $\geq$20\% for each camera and the
data should not be trusted within that region.  The
three LSB galaxies with detected CO emission are on the left, while three
LSB galaxies with no detectable molecular gas are on the right.
Note that the underlying shape of UGC 01922's spectra is likely due to
SN1987s which occurred approximately 40\acs\ from the galaxy's center \citep{oneil02b}.\label{fig:optical1}}
\end{figure}

\begin{figure}[h]
\plotone{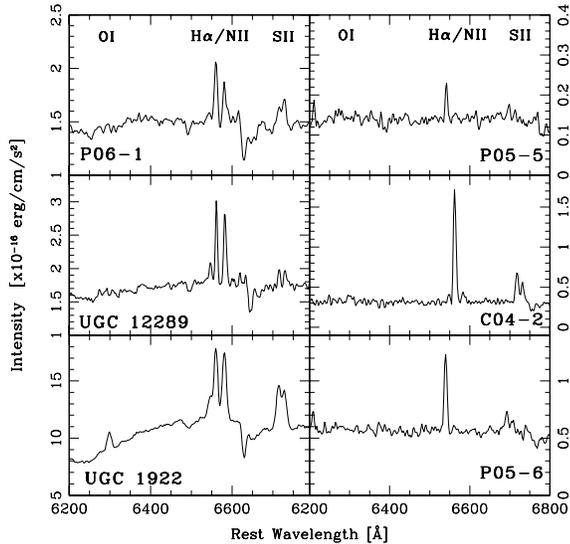}
\caption{Blow-up on the [\ion{O}{1}], [\ion{N}{2}], H$\alpha$, and [\ion{S}{2}] region
of the optical spectra.  As in Figure~\ref{fig:optical1}, the
three LSB galaxies with detected CO emission are on the left, while three
LSB galaxies with no detectable molecular gas are on the right.\label{fig:optical2}}
\end{figure}

The standard IRAF data reduction procedures as outlined in, e.g. \citet{vanz98}, were followed.
Each frame individually underwent bias subtraction and flatfielding.  Frames of the same
source and taken on the same night were median combined.  The resultant image then
underwent sky subtraction and wavelength and photometric calibration.  Images of the
same object taken at different times were then combined (using IRAF's SCOMBINE procedure),
and extinction and foreground reddenning due to our Galaxy was corrected.
Wavelength calibration was
obtained by observation of arc lamps taken before and after each galaxy observation.  A hollow
cathode (FeAr) lamp was used to calibrate the blue wavelengths, and a HeNeAr lamp was used
to calibrate the red camera.  
Flux calibration was obtained by observation of a minimum two standard stars each night
\citep{oke83,oke90,massey88}.  Extinction correction was done based off the
KPNO extinction table given for IRAF's ONEDSPEC routine.  Galactic extinction was corrected
for using the reddening law of \citet{seat79} as parameterized by \citet{howa83}.
Values for A$_B$ and E(B-V) are from \citet{schl98} and were obtained using NED
(the NASA Extragalactic Database){\footnote{NED is operated by the Jet Propulsion
Laboratory, California Institute of Technology, under contract with NASA.}}.

Errors were determined as follows.  Flux calibration errors, due primarily to differences
in the standard star calibrations, were consistently less than 10\%.  The r.m.s. errors
varied with wavelength across the spectra due to CCD response, calibration lamp accuracy,
etc.  Rms errors were determined through taking into account the Poisson noise of the
line, the error associated with the sensitivity function, read noise, and sky noise.
Note that no error was assumed for the effects of the assumed Balmer absorption
or the reddening coefficient. 

Results from the optical observations are shown in Figures~\ref{fig:optical1} and \ref{fig:optical2}
and Table~\ref{tab:optical_obs}.

\section{Results}

\subsection{CO Observations \label{sec:obs}}

The results of our CO observations are
given in Table~\ref{tab:COprops} and Figures~\ref{fig:codets} and
\ref{fig:conondets}. Of the seven galaxies observed for this project, 
CO was detected in only two -- UGC~01922 and UGC~12289.
Both detected galaxies have clear (5--6$\sigma$) detections
in the J(1--0) line and a $\ge$4$\sigma$ detection in the J(2--1) line.  After smoothing to
34 \kms\ channel$^{-1}$, the observed main beam
brightness temperature was 3.4$\pm$0.7/7.3$\pm$0.8 mK and
5.8$\pm$0.9/3.5$\pm$1.0 mK for the J(1--0)/J(2--1) lines
of UGC~01922 and UGC~12289, respectively.  The galaxies have
fluxes of
1.38$\pm$0.30/2.96$\pm$0.32 K \kms\ and 1.16$\pm$0.27/0.69$\pm$0.23
K \kms\ for UGC~01922 and UGC~12289, respectively.  (Errors for the flux are computed from the 
r.m.s. error of the main beam brightness temperature and an assumed 
34 \kms\ error for the velocity width.)

The 3$\sigma$ non-detection limits for the five galaxies without CO detections range
between 1.5 -- 3.6 mK.  These detection limits
are similar to those found by \citet{oneil00b}, and
are a factor of 6-7 times smaller than the CO studies on 
LSB galaxies done by both \citet{deblok98} and 
\citet{schom90}.  The integrated flux limits,
though, are only slightly smaller than those in
the \citet{deblok98} study due to smaller total
mass (narrower velocity widths) of their sample.

The apparent optical diameters of the galaxies in this study range
from 0.4$^\prime$ -- 2.8$^\prime$ (Table~\ref{tab:props}),
while the beam size of the IRAM telescope is 11\acs\ and
22\acs\ for the J(2--1) and J(1--0) lines, respectively.
Using 2.6-mm CO observations of more than 200 high
surface brightness (HSB) spiral galaxies, \cite{young95}
found the average extent of CO gas in the studied galaxies
are one-half the optical radius.  If this trend holds
true for LSB galaxies, it would indicate the IRAM beam enclosed
the entire (potential) reservoir of CO gas only
for \obc~N09-2, while the other
galaxies had 50\% (UGC~12289, \obc~P07-1, \obc~P05-6)
and 35\% (UGC~01922, UGC~06968) of their potential
CO radius enclosed by the J(1--0) beam.
However, a recently obtained synthesis map of the CO J(1$-$0)
emission of UGC 01922 obtained with the Owens Valley Radio
Observatory millimeter-wavelength array shows Young, et.al's findings
may not be applicable to LSB systems.  Approximately 65\%
of the flux found with the IRAM 30m telescope was detected at OVRO
within a $\sim$10\acs\ disk. This shows that at least in UGC 01922
the CO emission is dominated by a brighter
compact nuclear region (which would readily fall within the
IRAM beam), rather than consisting solely of a large-scale diffuse
(low surface density) disk.

The integrated CO flux (or upper limits) of the observed
galaxies were converted into an H$_2$ mass by assuming
a value of $\rm N(H_2)/\int{T(CO)dv}\;=\;3.6\;\times\;10^{20}
\;cm^{-2}/(K\;km\;s^{-1})$ adopted from \citet{sanders86}. 
It has often been argued that the CO $\leftrightarrow$ H$_2$
conversion factor (X) varies with ISM density structure, metallicity and
\ion{H}{1} column density \citep{mihos99, isreal97, wilson95, maloney98}. 
If this is the case, then the X factor used in our calculations
could be low by as much as a factor of 10 \citep{mihos99}.
Regardless, to allow for a ready comparison between the objects,
the X factor used for our data analysis is the 
same as that used in both all previous LSB galaxy CO studies 
as well as in all other CO studies discussed in this paper.
Errors due to this assumption are discussed later in this paper
(Section~\ref{sec:sf}).

\subsection{Optical Spectra}

Looking at both Table~\ref{tab:optical_obs} and Figure~\ref{fig:optical1},
the spectra appear to fall into two categories -- systems containing  both a 
significant number of absorption lines as well as high [\ion{N}{2}/H$\alpha$]
ratios, and those spectra devoid of absorption lines and containing
[\ion{N}{2}/H$\alpha$] ratios typical of a younger stellar population.
Significantly, this breaking of the optical spectra into two distinct groups
coincides with the detection (or lack thereof) of molecular gas in the
galaxies.  That is, the three galaxies which have detectable amounts of
CO also have optical spectra indicative of an older stellar population
with a potential AGN/LINER core, while the galaxies without any discernable
molecular gas appear to have a young overall stellar age with no indication
of any high-energy star-formation activity within the galaxies' cores.
Note that the underlying shape of UGC 01922's spectra is likely due to
SN1987s which occurred approximately 40\acs\ from the galaxy's center \citep{oneil02b}.

\section{Comparison with Other Galaxies}

Combining all CO observations taken of LSB galaxies provides us with
a compendium of 34 galaxies, three of which have been detected
at both CO J(1--0) and J(2--1).  Although this is by no means a 
statistically complete sample, it is enough to
at least try and understand the molecular gas content
within these enigmatic systems.

\subsection{Finding CO in LSB Galaxies}

\begin{figure}[t]
\plottwo{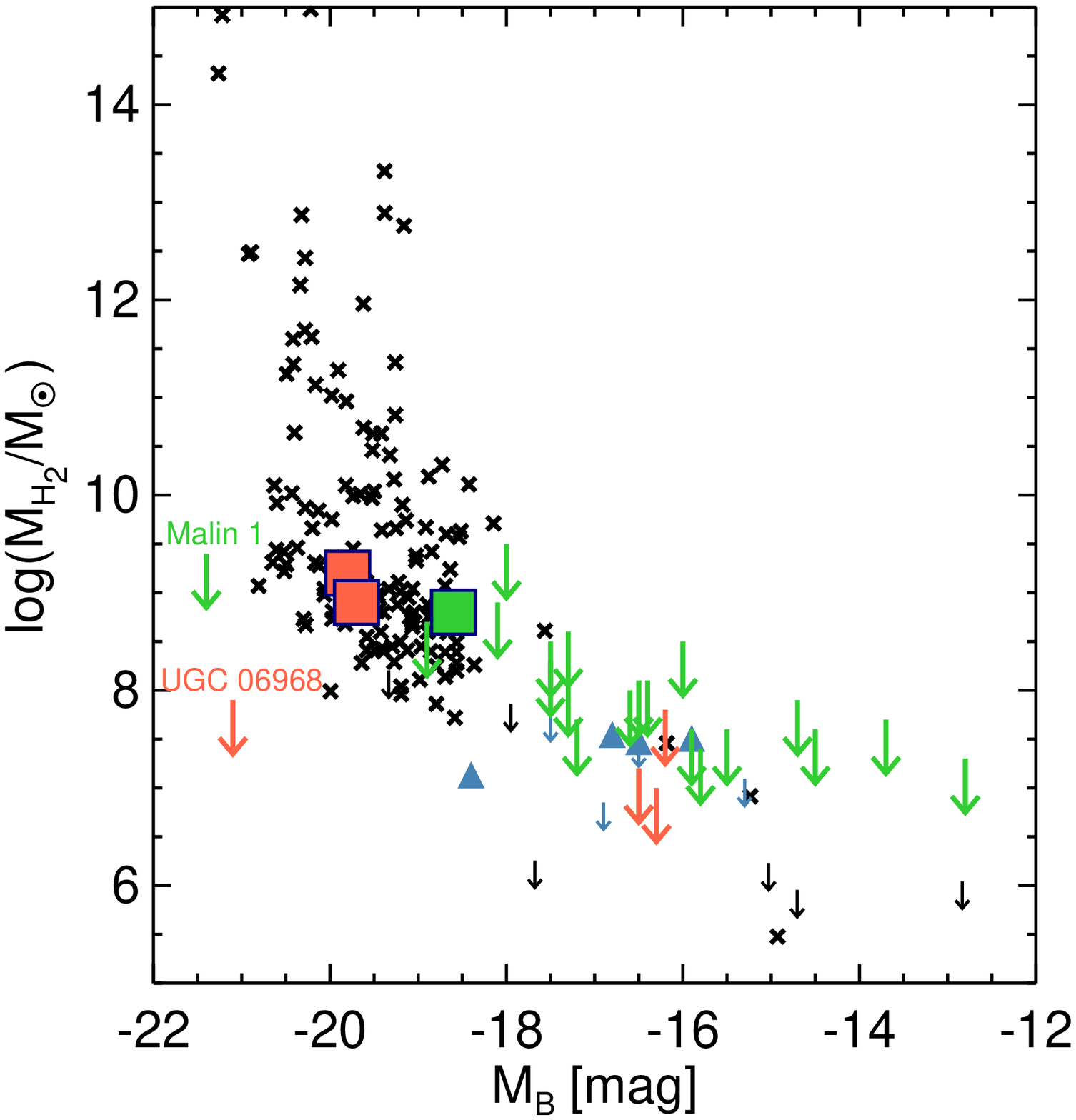}{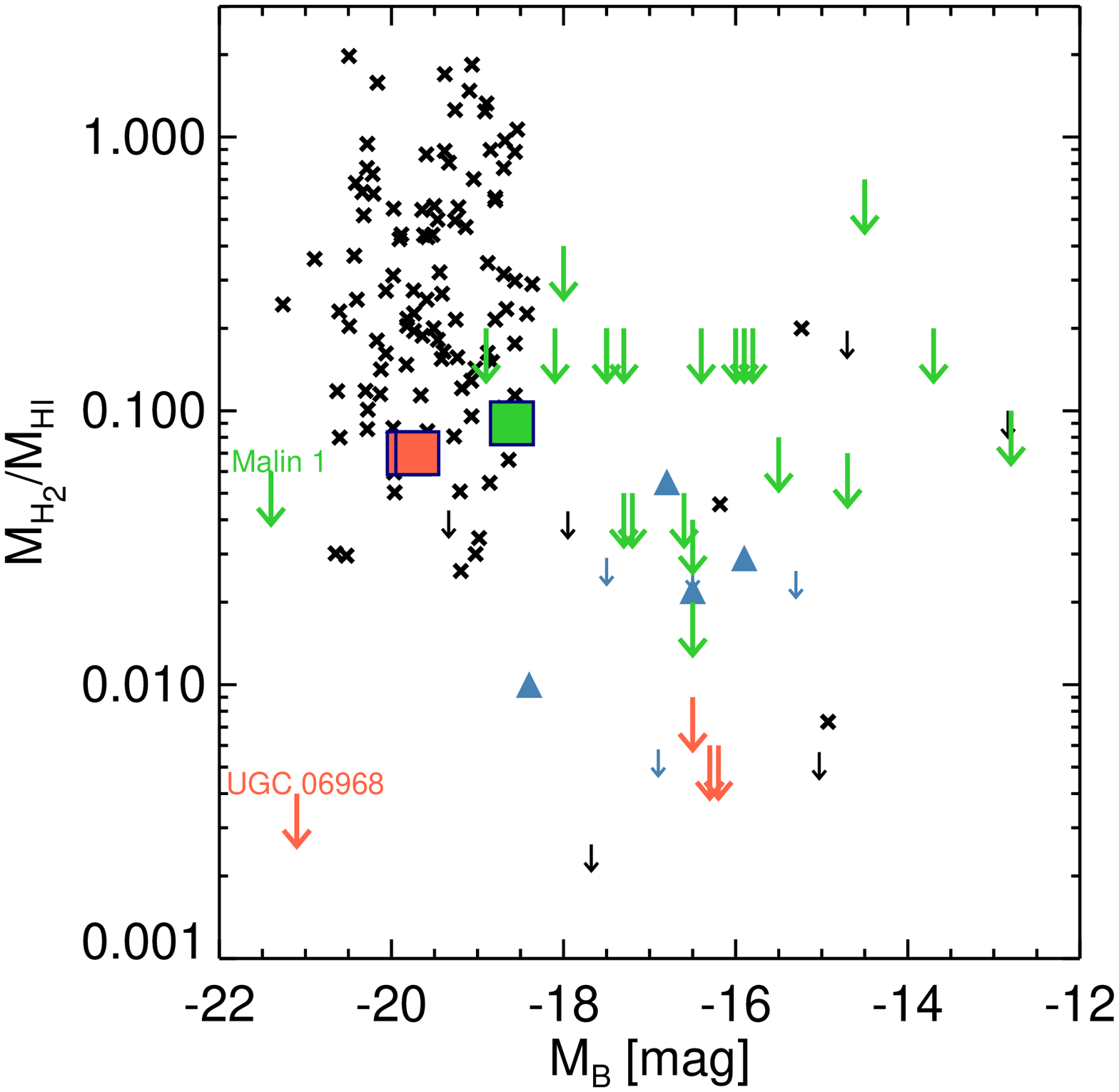}
\caption{Absolute (B) magnitude versus H$_2$ mass (left) and the H$_2$-to-\ion{H}{1} mass
ratio (right).  The red symbols are LSB galaxies from this survey, the
green symbols are LSB galaxy measurements from previous surveys \citep{oneil02, 
braine00, deblok98, schom90}, the blue symbols are from the \citet{matthews01}
study of CO in extreme late-type spiral galaxies, and 
the black symbols are taken from various studies of the CO content in 
HSB spiral galaxies \citep{casoli96, boselli96, tacconi87}. Note that 
the two LSB galaxy detections from this survey are overlapping in the
figure on the right.  An arrow indicates only an upper limit was found.
Both Malin 1 and UGC 06968, the two massive LSB galaxies discussed in Section 5.3,
are labelled on these plots.\label{fig:MB}}
\end{figure}

\begin{figure}[t]
\plottwo{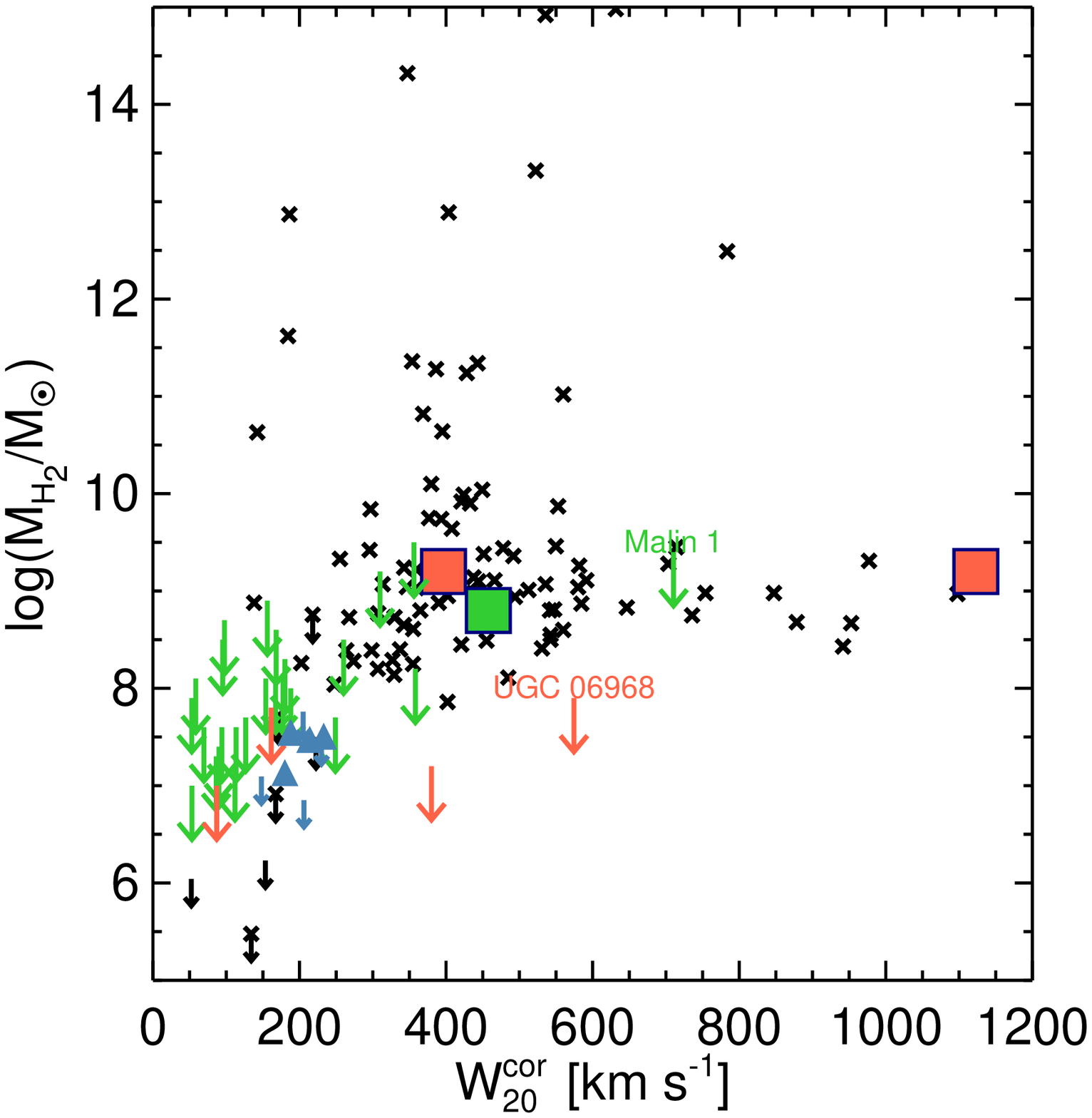}{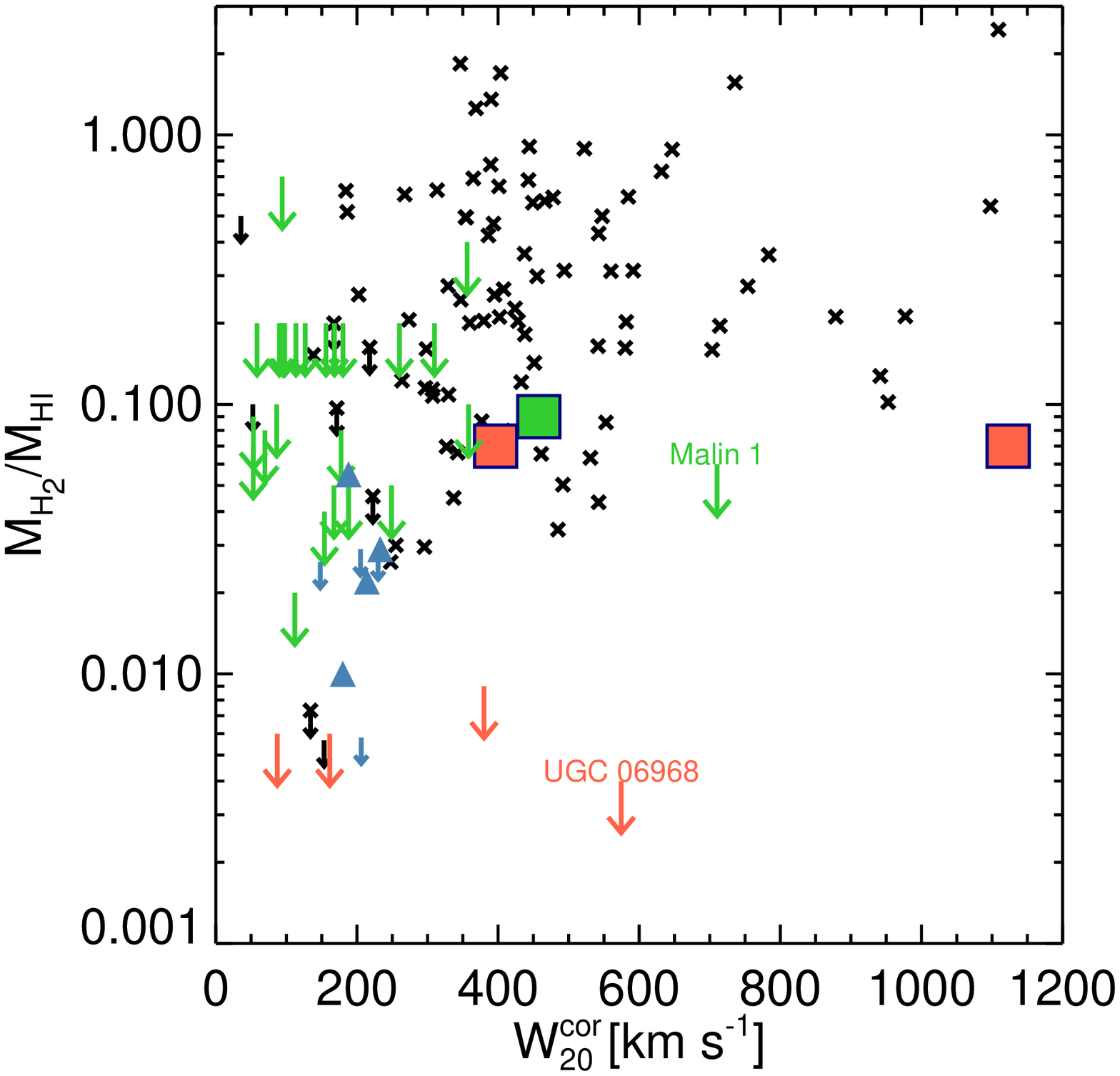}
\caption{Inclination corrected \ion{H}{1} velocity widths versus
H$_2$ mass (left) and the H$_2$-to-\ion{H}{1} mass
ratio (right).  The red symbols are LSB galaxies from this survey, the
green symbols are LSB galaxy measurements from previous surveys \citep{oneil02,
braine00, deblok98, schom90}, the blue symbols are from the \citet{matthews01}
study of CO in extreme late-type spiral galaxies, and
the black symbols are taken from various studies of the CO content in
HSB spiral galaxies \citep{casoli96, boselli96, tacconi87}.  
An arrow indicates only an upper limit was found.
Both Malin 1 and UGC 06968, the two massive LSB galaxies discussed in Section 5.3,
are labelled on these plots.\label{fig:v20}}
\end{figure}

In Tables~\ref{tab:props} and \ref{tab:COprops} we compile a list of the global
properties of all LSB galaxies which have (published)
CO observations.
Looking through the tables it can be seen that 
the galaxies with CO detections (``CO LSB galaxies") are not distinguished from the
non-detections (``non-CO LSB galaxies") 
by their morphological type, central surface brightness,
inclinations, or color (in the one case of a CO LSB galaxies with a measured $B-V$ color).
The only distinguishing feature among the CO LSB galaxies
is an apparent higher-than-average size and mass.  That is, the
absolute magnitudes, \ion{H}{1} gas mass, optical diameter,
and total mass (as defined by the inclination corrected
velocity widths) are on average considerably higher 
for the CO LSB galaxies than for the non-CO LSB galaxies
(Figures~\ref{fig:MB} \& \ref{fig:v20}).

A second factor which sets the CO LSB galaxies 
apart the majority of the non-detections is the presence of an
extremely active central region (e.g. LINER/AGN).
Eight LSB galaxies have both CO
observations and optical spectra taken across the galaxy's 
core.  (The two optical spectra not given in this
paper can be found in \citet{schom98} -- UGC 06968, and \citet{braine00}
-- Malin~1.)  Out of these, all three of the CO LSB galaxies 
have high \ion{N}{2}/H$\alpha$ line ratios,
while only two of the five non-CO LSB galaxies (UGC 06968 and Malin~1)
show this.  Additionally, both of the CO LSB galaxies
covered by 2MASS\footnote{2MASS, the Two Micron All Sky Survey is a
joint project of the University of Massachusetts and the Infrared Processing
and Analysis Center/California Institute of Technology, funded by the National
Aeronautics and Space Administration and the National Science Foundation.}
have entries in the 2MASS extended source catalog (UGC~01922: J=11.60$\pm$0.02,
H=10.80$\pm$0.02, K$_s$=10.47$\pm$0.02; UGC~12289: J=13.30$\pm$0.05,
H=12.60$\pm$0.06, K$_s$=12,24$\pm$0.07), while only two of the remaining
27 non-CO LSB galaxies covered by published 2MASS data
are in the catalog (UGC~06968: J=12.45$\pm$0.02,
H=11.69$\pm$0.02, K$_s$=11.50$\pm$0.03; Malin~1: J=14.54$\pm$0.06,
H=13.83$\pm$0.08, K$_s$=13.7$\pm$0.1).  Combined, these results
sugggest the CO LSB galaxies have more active nuclear regions than the
non-CO LSB galaxies.

LSB galaxies with detectable quantities of CO appear to have
both high masses/sizes and more active nuclear regions than
the non-CO LSB systems.  Yet even the combination of these
two quantities does not guarantee that an LSB system will
be detectable in CO.  Both UGC~06968 and (of course) Malin~1
have velocity widths and absolute magnitudes to rival those
of the three CO LSB galaxies and similar optical spectra, yet neither of these two
galaxies have detectable quantities of CO to very low limits
($\rm \int_{CO(1-0)}{T_{MB}dv}\;<$0.21, 0.15 K \kms\ and
$\rm log({M_{H_2}/M_\odot})\;<$7.9, 9.4 for UGC~06968 
and Malin~1, respectively).  As a result, although it would
appear that having a high mass and active nucleus raises the
probability of an LSB galaxy containing a detectable amount 
of CO, high mass by no means guarantees a galaxy will 
have sufficient CO content for detection.

\subsection{Comparing the Molecular Gas Content of LSB and HSB Galaxies}

\begin{figure}[h]
\plottwo{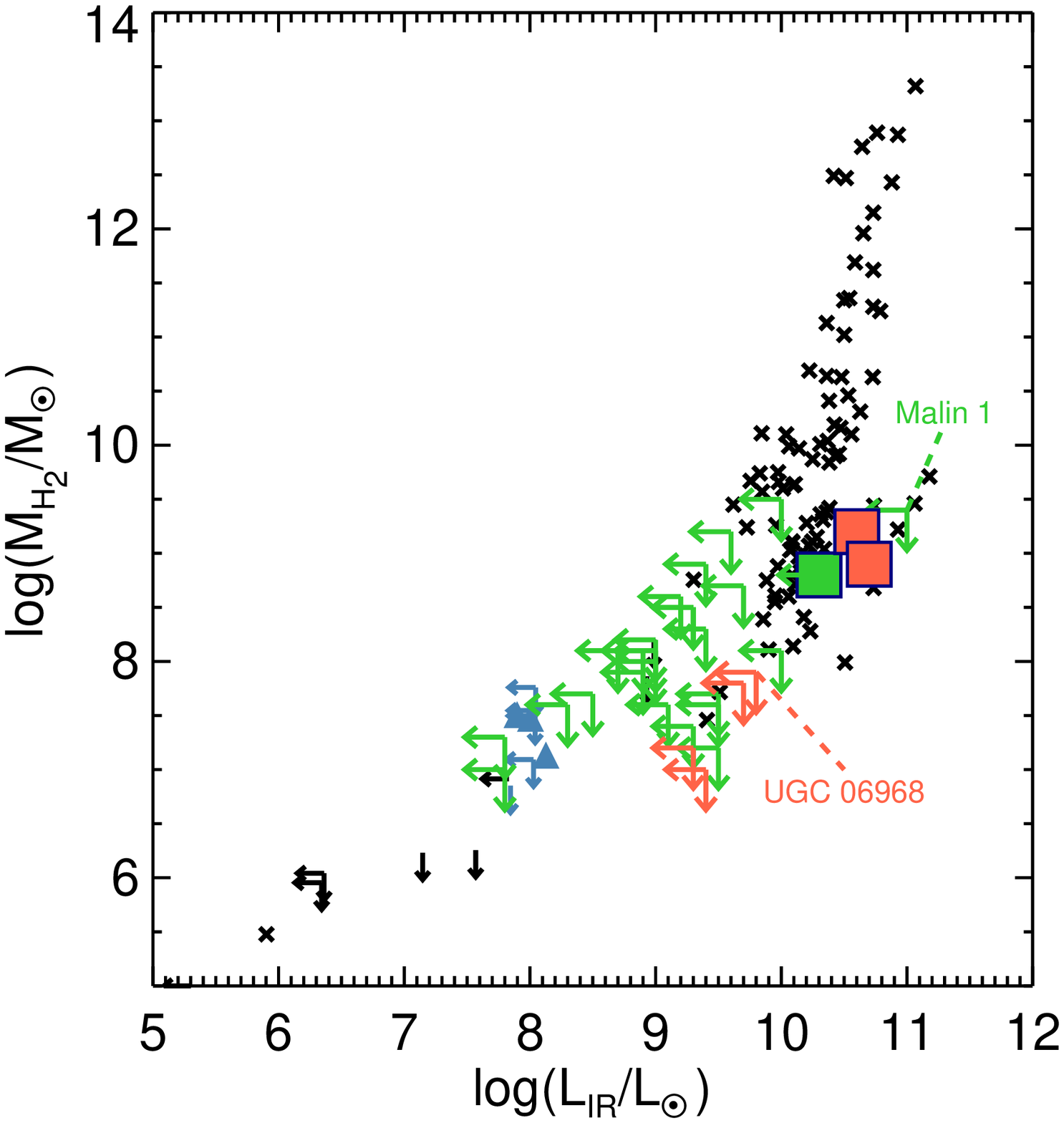}{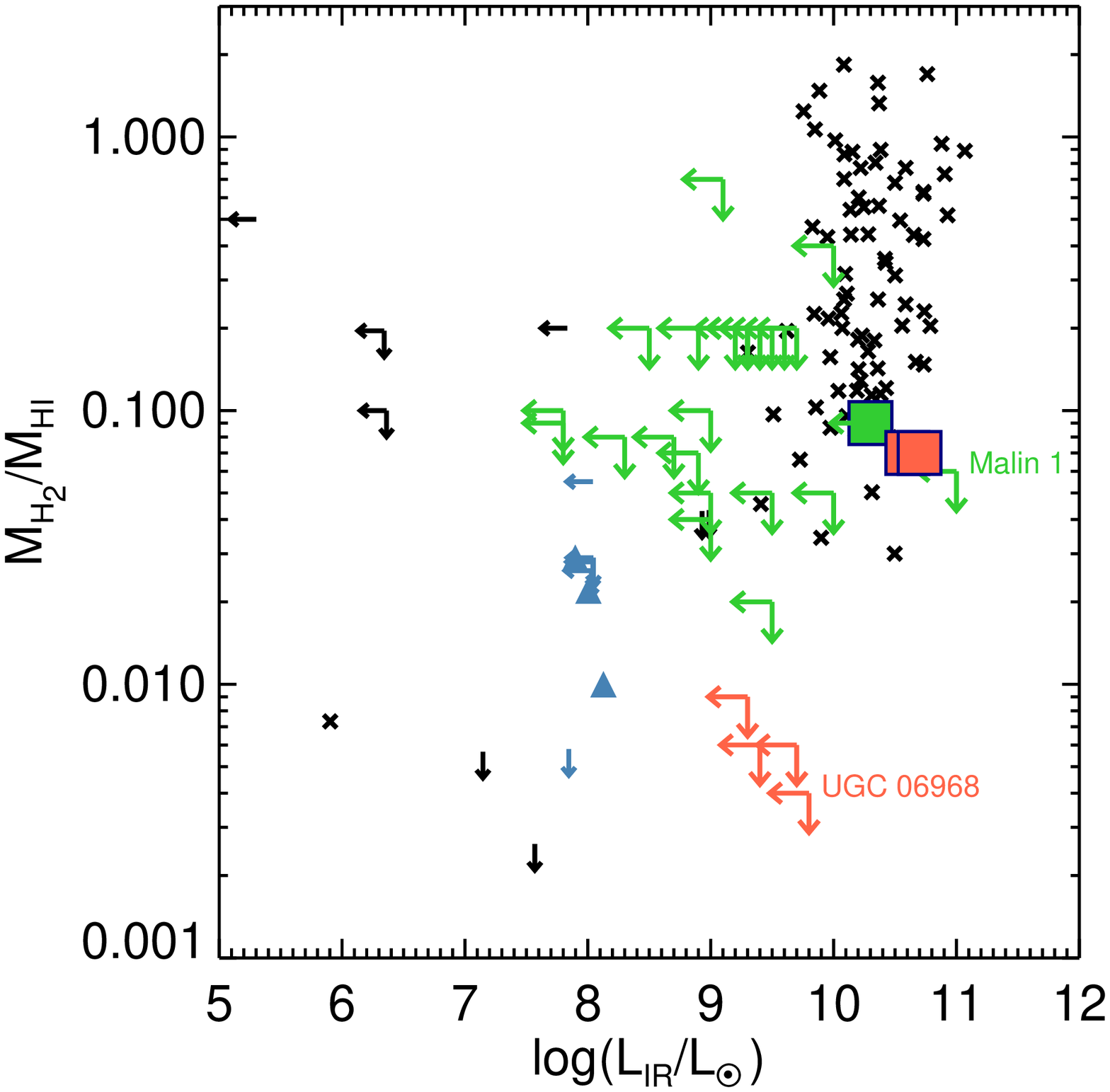}
\caption{Infrared luminosity versus
H$_2$ mass (left) and the H$_2$-to-\ion{H}{1} mass
ratio (right).  The red symbols are LSB galaxies from this survey, the
green symbols are LSB galaxy measurements from previous surveys \citep{oneil02,
braine00, deblok98, schom90}, the blue symbols are from the \citet{matthews01}
study of CO in extreme late-type spiral galaxies, and
the black symbols are taken from various studies of the CO content in
HSB spiral galaxies \citep{casoli96, boselli96, tacconi87}.  
An arrow indicates only an upper limit was found.  Note that
only the upper limits of L$_{IR}$ were found for the LSB galaxies.
Both Malin 1 and UGC 06968, the two massive LSB galaxies discussed in Section 5.3,
are labelled on these plots.\label{fig:FIR}}
\end{figure}

\begin{figure}[h]
\plotone{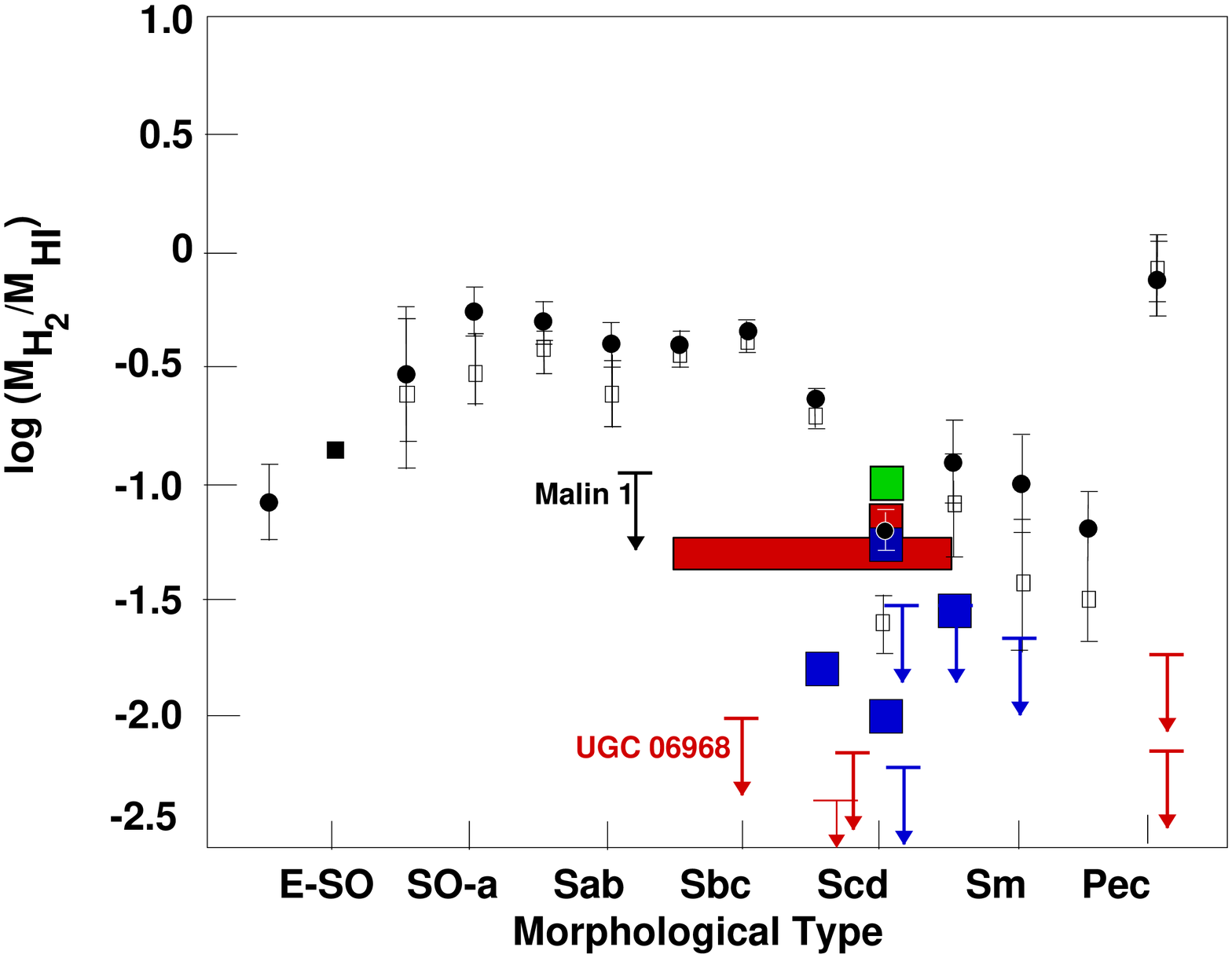}
\caption{Variation of the molecular-to-atomic gas fraction along the morphological type sequence.
The red symbols are the individual LSB galaxies in this survey,
the green symbol is the one LSB galaxy detection from a previous
survey \citep{oneil00b}.
The size of the boxes represent the errors
associated with our observations.  The black symbols are taken from
\citet{casoli98}, and are the mean values taken from a compendium of 582 CO
observations of disk galaxies.  The filled circles show mean values 
with upper limits treated as detections, and open squares give the mean
values when upper limits are taken into account using a
survival analysis method. The error bars in these points give the error
on the mean.   For comparison, the blue symbols are the individual extreme
late-type galaxy sample of \citet{matthews01}.
Both Malin 1 and UGC 06968, the two massive LSB galaxies discussed in Section 5.3,
are labelled on these plots.
\label{fig:casoli_fig4}}
\end{figure}

Figures~\ref{fig:MB} -- \ref{fig:casoli_fig4} compare the 
results from this and all other LSB galaxy CO studies
with a sample of measurements from a variety of other galaxy 
studies.  These include `standard' HSB disk galaxy studies
\citep{boselli96,casoli96}, dwarf galaxy studies \citep{tacconi87},
and a study of extreme late-type spiral galaxies \citep{matthews01}.
In all cases a conversion factor of $\rm N(H_2)/\int{T(CO)dv}\;=
\;3.6\;\times\;10^{20}\;cm^{-2}/(K\;km\;s^{-1})$ was used to
allow ready comparison between the results.

In all the CO non-detections in our survey, the upper limits placed on
M$_{H_2}$/L$_B$ are significantly lower than the upper limits found in previous
LSB galaxy CO studies.  However,
examining the various plots (Figures~\ref{fig:MB} -- \ref{fig:casoli_fig4})
one can see that in the majority
of cases the upper limits placed on M$_{H_2}$ and M$_{H_2}$/M$_{HI}$
in LSB galaxies are not unreasonably low when compared to the trends seen in
the majority of other spiral galaxy studies.  That is, although the LSB galaxies
typically lie at the low M$_{H_2}$ and M$_{H_2}$/M$_{HI}$ end of the spectrum, the
range of the data seen in the plots of
M$_{H_2}$ and M$_{H_2}$/M$_{HI}$ versus M$_B$, $W_{20}^{cor}$,
L$_{FIR}$ and morphological type for the HSB and dwarf galaxies
covers the data points from the majority of the LSB galaxies.  The three
CO LSB galaxies also have M$_{H_2}$ and M$_{H_2}$/M$_{HI}$ values
within the HSB galaxy range, but again lying at the low edge of the 
M$_{H_2}$ and M$_{H_2}$/M$_{HI}$ distribution as defined by HSB galaxies.

There are, though, two exceptions to this rule --  UGC~06968 and Malin~1.
Both have upper limits on M$_{H_2}$ and M$_{H_2}$/\Msol\ 
well below that of other galaxies with the
same velocity width, magnitude, and/or morphological type (Figures~\ref{fig:MB},
\ref{fig:v20}, and \ref{fig:casoli_fig4}).  It is interesting to note that
these galaxies are the only two LSB systems with masses and
luminosities rivaling that of the three LSB galaxies with CO detections,
as well as having similar optical emission lines and near-IR emission.
These galaxies are further discussed in Section~\ref{sec:gals}.

\section{Star Formation and Molecular Gas in LSB Systems \label{sec:sf}}

\subsection{Theories on the Efficacy of Forming Molecular Gas in LSB Galaxies \label{sec:efficacy}}

The previous lack of CO detection in LSB galaxies has been
attributed to a number of different factors. One argument used
to explain the low molecular gas content in LSB galaxies 
springs from the low metallicities commonly associated with
LSB disks \citep{deblok98,mcgaugh94b}.   Low metallicities mean there will be a lack
of dust grains to drive molecule formation, resulting in
a low molecular-to-atomic gas ratio.  Additionally, the
dust grains provide shielding against the interstellar
radiation field for the H$_2$ and CO molecules, again
reducing the amount of molecular gas within the galaxies.
Finally, as the CO molecule more readily photodissociates,
the lack of dust grains will result in a lower CO-to-H$_2$
ratio and thus a higher CO-to-H$_2$ conversion factor
\citep{oneil00b,mihos99,deblok98,schom90}.
 
An alternative scenario used to explain the low CO fluxes
found for LSB galaxies is to impose a truncated initial mass
function (IMF) on a constant star formation rate (SFR) within
LSB systems \citep{schom90}.  This SFR could be adjusted to match the 
estimated amount of H$_2$ seen in LSB galaxies.  Additionally, the IMF 
would have to be skewed to produce a high fraction of A and F stars 
and relatively few O and B stars.  This would result in a reduction
of the heavy element yield during the star formation process,
lowering the overall molecular gas content of LSB galaxies.

\citet{braine00} offered the argument that, at least within the
environment of Malin~1 (and presumably similar LSB galaxies),
the effects of metallicity on the CO-to-H$_2$ conversion factor are
not enough to raise the molecular gas content to the amount expected
based off the predicted SFR.  Instead, they  argue that the gas (and 
dust) in Malin~1 is so cold as to reduce the CO emission from molecular
clouds to a rate much less than that of our galactic disk.  This would
considerably reduce the CO emission per H$_2$ mass and, when combined 
with the metallicity effects, could be used to explain the low
upper limit placed on the CO flux.

A final argument often invoked to explain the previous lack
of CO detection within LSB galaxies results from the low
baryonic column densities found within these systems  \citep{oneil00b,mihos99,deblok98,schom90}.
First, the low dust column densities will, as mentioned earlier, 
result in a deficit of dust grains to act as a catalyst in
the conversion of \ion{H}{1} into H$_2$.  Secondly, the diffuse
ISM will result in less shielding against UV photons.  
Lastly, the low column densities result in a low overall
SFR, reducing the metallicity of the system, with the effects
mentioned above.

\subsection{Comparing the Results with the Theories}

For the first time we have available a catalog of over 30 CO
measurements of LSB galaxies, including three detections.
As a result we can compare the properties of the CO 
and non-CO LSB galaxies with the various theories
given in the last section with the hope
of understanding better the processes which govern
molecular gas within LSB systems.

In Section 4.2 we determined that the factors which 
differentiate the CO LSB galaxies and the majority of the
non-CO LSB galaxies are the galaxies' sizes, masses, and
central bulges.  (The two notable exceptions to this
rule, UGC~06968 and Malin~1, will be discussed in the next
subsection.)  It seems likely that the presence of
a central bulge within LSB galaxies is directly tied
to the mass of the systems -- the more massive a galaxy
is, the larger its gravitational potential, and the more
likely it will have a central mass concentration.  This
supposition is borne out with observation, as the majority 
of LSB galaxies known with M$_B<-$19 have bulges
\citep{sprayberry95, pickering97}.  The fact that the
more massive LSB systems have a centrally concentrated
region of high baryonic density means that within these galaxies' core
the majority of arguments governing
low CO and/or molecular gas content in the low density
environment of LSB disks do not hold.  Instead the
central regions of these galaxies should undergo a similar
star formation history to the core of a `typical' HSB 
spiral galaxy, albeit with a slower gas infall rate.  
This idea is confirmed by the optical spectra we obtained
of the three LSB galaxies with and without detected CO emission,
as the three CO LSB galaxies have optical spectra indicative of
a much older stellar population than is found for the
three non-CO galaxies. 

However, even including the three LSB galaxies with CO detections,
it is clear that the CO content of LSB galaxies is lower 
than that found in an `average' HSB spiral galaxy.
That is, although both the measured values and
upper limits show the CO content of LSB galaxies lie within
the range set by HSB galaxies, the LSB systems lie on the
low end of the HSB galaxy distribution range, but 
not so low as to set the LSB galaxies apart.  
Going through the possibilities laid out in Section~\ref{sec:efficacy},
it seems clear that the explanation for the lower CO fluxes found in
LSB systems is directly related to the overall lower
density intrinsic to LSBs.  Within the pure disk LSB
galaxies it is probable that the low upper limits placed
on the CO flux is due both to a lower than average  
molecular gas content and to a CO-to-H$_2$ conversion
factor which is higher than that found within, e.g.
the Milky Way.  Similarly, the lower CO fluxes found within
the bulges of the more massive LSB galaxies are 
likely due to an low infall rate of gas feeding the star formation
processes within the bulge, and a lower metal content
within the infalling gas.

\subsection{UGC~06968 and Malin~1 \label{sec:gals}}

UGC~06968 and Malin~1 are remarkable for being two of the 
largest and most massive of the LSB galaxies studied.
They are the only two LSB galaxies studied with M$_B\;<\;-$21
(the usual cut-off for classifying an LSB galaxy as a massive system --
see, e.g. Pickering, et.al 1997).
Additionally, they are two of only four galaxies with 
M$_{HI}\;>\;10^{10}$~\Msol\ and $\rm W_{20_{cor}}^{HI}\;>$ 550~\kms.
As it appears that only the massive LSB galaxies have detectable
quantities of CO, it is worthwhile to ask the question of why,
when CO was readily detected in the other three massive LSB systems,
no molecular gas has been seen in either UGC~06968 or Malin~1. 

Because of its distance (v$_{HEL}$ = 24733 \kms), the upper limits
placed on Malin~1's absolute CO flux measurement are not exceptionally 
low -- $\rm M_{H_2} < 10^{9.4}$~\Msol\ and $\rm M_{H_2}/M_{HI} <$~0.06.
Regardless, when compared with the H$_2$ masses found for other HSB and LSB
galaxies of similar mass, Malin~1's upper limits indicate it has a molecular
gas content below that found for all other galaxies of its
size, mass, or magnitude.  Additionally, Malin~1 has a surprisingly low
value for L$_{FIR}$ as measured by the IRAS upper limits (Table~\ref{tab:props}).
\citet{braine00} state that, based primarily off the low L$_{FIR}$ values, the gas and dust in
Malin~1 is quite cold, reducing the galaxy's CO-to-H$_2$ fraction.
From this they conclude that the H$_2$ mass of Malin~1 is likely 
near that of similar spiral galaxies, and it is only the CO gas
which is lacking. 

Classified as an AGN/LINER system by \citet{schom98}, UGC~06968
falls well off the expected range of values for both $\rm M_{H_2}$
and $\rm M_{H_2}/M_{HI}$.  Looking at Figures~\ref{fig:MB} and \ref{fig:v20}
we can see that the upper limit found for UGC~06968's H$_2$ mass
is at least $\sim$10 times less than would be expected, while the
galaxy's molecular-to-atomic mass ratio ($\rm M_{H_2}/M_{HI}$)
is only 5\% that of two LSB galaxies with CO detections and
similar global properties.  Like Malin~1, which has a similar morphology
and classification, UGC~06968 has a low
L$_{FIR}$ limit for its size, indicating it, too, may contain
extremely cold gas and dust, reducing its CO flux and increasing
its H$_2$-to-CO conversion factor.  Unlike Malin~1, though, 
the upper limits placed on UGC~06968's CO flux are low
enough that a low gas temperature alone cannot increase
the galaxy's H$_2$-to-CO conversion factor enough to raise $\rm M_{H_2}$
up to the value found for similar HSB systems.    Looking at the
various models of \citet{mihos99}, the clearest means of increasing
UGC~06968's H$_2$ mass to the lowest value found for any other
galaxy with similar global properties is by reducing the metallicity
of the galaxy's {\it bulge} to something near 1/5 the
solar value.   (In the Mihos, et.al models, the lower metallicities result
in fewer C and O atoms from which to form CO.  This increases the CO-to-H$_2$
conversion factor in a manner analagous to that found by, e.g. Wilson 1995, Israel 1997,
and Maloney \& Black 1988.)  Even if this is done, though, the $\rm M_{H_2}/M_{HI}$
value for UGC~06968 will be half that found for galaxies with
similar masses and luminosities.  A second possibility is that, unlike 
UGC 01922, the CO gas within UGC 06968 may not be concentrated within
its core but instead may be spread throughout the galaxy's disk.
As the disk of UGC 06968 extends well beyond that of the IRAM beam,
this would imply we observed only a small fraction of UGC 06968's
total molecular gas.  If this is the case, though, it still would not
explain why UGC 06968's molecular gas distribution is so different
from other galaxies with similar morphologies.  As a result,
it is more likely that our observations did include the majority of the
molecular gas within the galaxy, and UGC 06968 is
simply deficient in CO gas.

\section{Conclusion}

We have obtained seven new, deep CO observations of LSB galaxies.
Combined with all previous CO observations taken of LSB systems,
we have a total of 34 observations, in which only 3 galaxies have
had detections of their molecular gas.  Looking at the differences
between both the CO and non-CO LSB galaxies and
and a collection of HSB galaxies with CO observations
indicates it is primarily the lower baryonic density intrinsic to LSB galaxies
which is causing their low CO fluxes.  As has been discussed in
previous papers, this in turn is likely responsible both for
a low CO-to-H$_2$ ratio within the galaxies and for a lower
density of molecular gas within LSBs.  As a result, it is 
only those LSB galaxies which contain regions of higher
baryonic density, such as the high mass LSB galaxies with
distinct central bulges, in which CO gas will be readily detectable.

It is interesting to note that two of the most massive LSB galaxies
studied have upper limits placed on their CO fluxes well below that
of similar LSB and HSB galaxies.  In the case of Malin~1 the difference
between the upper limits placed on the galaxy' CO gas can be largely 
explained through assuming a very low temperature for the galaxy's
gas and dust \citep{braine00}.  In the case of UGC 06968, though,
temperature alone cannot be used to explain the galaxy's low
M$_{H_2}$ and extremely low M$_{H_2}$/M$_{HI}$ values.  Instead
we must resort to also assuming a very low metallicity within
the galaxy's core and/or a much lower baryonic density even in
the galaxy's central region than its optical morphology and
AGN/LINER characteristics suggest.

Even though the majority of observations taken of LSB galaxies
result in only upper limits on the galaxies' CO flux, we are gradually
gaining a clearer picture of the ISM, and consequently the SFH of these
enigmatic systems.  As the theories and observations are continued, we should 
gain a much clearer understanding of the star formation processes within
the low density environments of LSB galaxies.

\acknowledgements{Thanks to Greg Bothun for providing the PMO images of
UGC 12289 and to Liese van Zee for help with taking and reducing the 
Palomar spectra.  P.H. acknowledges partial support from the Puerto Rico
Space Grant Consortium and from NSF grant EPS-9874782.
This research has made use of the NASA/IPAC Extragalactic Database (NED) which is operated by the Jet
Propulsion Laboratory, California Institute of Technology, under contract with the National Aeronautics and Space
Administration. }

\clearpage

\begin{thebibliography}{}
\bibitem [Bergvall, et.al(1999)]{bergvall99}
Bergvall, Nils, Rvnnback, Jari, Masegosa, Josefa, \& Vstlin, Gvran 1999 A\&A 341, 697

\bibitem [de Blok \& van der Hulst(1998)]{deblok98}
de Blok, W.J.G, \& van der Hulst, J.W. 1998 A\&A 336, 49D

\bibitem [de Blok \& McGaugh(1997)]{deblok97}
de Blok, W.J.G, \& McGaugh, S. 1997 MNRAS 290, 533

\bibitem [de Blok, McGaugh, \& van der Hulst(1996)] {deblok96}
de Blok, W.J.G, McGaugh, S., \& van der Hulst, J.W. 1996 MNRAS 283, 18

\bibitem [de Blok, van der Hulst \& Bothun(1995)] {deblok95}
de Blok, W.J.G, van der Hulst, J.W., \& Bothun, G. 1995 MNRAS 274, 235

\bibitem[Boselli, et.al(1996)]{boselli96}
Boselli, A., Mendes de Oliveira, C., Balkowski, C., Cayatte, V., Casoli, F. 1996 A\&A 314, 738

\bibitem [Boselli \& Gavazzi(1994)] {bose94}
Boselli, A. \& Gavazzi, G. 1994 A\&A 283, 12

%
\bibitem [Braine, Herpin, \& Radford(2000)] {braine00}
Braine, J., Herpin, F., \& Radford, S. J. E. 2000 A\&A 358 494

\bibitem[Burkholder, Impey, \& Sprayberry(2001)]{burkholder01}
Burkholder, V., Impey, C., \& Sprayberry, D. 2001 AJ 122, 2318

(Buisson, et.al 2002).
\bibitem[Buisson, et.al(2002)]{buisson02}
Buisson, G., et.al (2002) {\it CLASS Continuum and Line Analysis System Handbook},
online at http://iram.fr/GS/class/class.html

\bibitem[Casoli, et.al(1998)] {casoli98}
Casoli, F., Sauty, S., Gerin, M., Boselli, A., Foqu\'e, P., Braine, J.,
Gavazzi, G., Lequeux , J., \& Dickey, J. 1998 A\&A 331, 451

\bibitem[Casoli, et.al(1996)]{casoli96}
Casoli, F., Dickey, J., Kazes, I., Boselli, A., Gavazzi, G., Jore, K. 1996 A\&AS 116, 193

\bibitem[Davies, Phillipps, \& Disney(1990)]{davies90}
Davies, J. I., Phillipps, S., \& Disney, M. J. 1990 MNRAS 244, 385

\bibitem [Eder \& Schombert(2000)] {eder00}
Eder, J. \& Schombert, J. 2000 ApJS 131, 47

\bibitem [Garnier, et.al(1996)] {garnier96}
Garnier, R.  Paturel, G., Petit, C., Marthinet, M. C., \& Rousseau, J. 1996 A\&AS 117, 467

\bibitem [Gavazzi(1987)] {gavazzi87}
Gavazzi, G. 1987 ApJ 320, 96

%
\bibitem [Giovanelli \& Haynes(1985)] {giov85}
Giovanelli, R. \& Haynes, M. 1985 AJ 90, 2445

%
\bibitem[Howarth(1983)]{howa83}
Howarth, I. 1983 MNRAS 203, 801

\bibitem[Huchtmeier, Karachentsev, Karachentseva(2000)] {huch00}
Huchtmeier, W.K., Karachentsev, I.D., \& Karachentseva, V.E. 2000 A\&AS 147, 187

\bibitem[van der Hulst, et.al(1993)]{vdhulst93}
van der Hulst, J. M., Skillman, E. D., Smith, T. R., Bothun, G. D., McGaugh, S. S.,
\& de Blok, W. J. G.  1993 AJ 106, 54

\bibitem[Impey \& Bothun(1989)] {impey89}
Impey, C. \& Bothun, G. 1989 ApJ 341, 89

\bibitem[Isreal(1997)] {isreal97}
Isreal, F. 1997 A\&A 328, 471

\bibitem[Kutner \& Ulich(1981)] {kut81}
Kutner, M. L. \& Ulich B. L.  1981, ApJ 250 341

\bibitem[Maloney \& Black(1998)] {maloney98}
Maloney, P., \& Black, J. H. 1988, ApJ 325, 389

\bibitem[Massey, et.al(1988)]{massey88}
Massey, Philip, Strobel, Kevin, Barnes, Jeannette V., \& Anderson, Edwin 1988 ApJ 328, 315

\bibitem[Matthews \& Gao(2001)] {matthews01}
Matthews, L. \& Gao, Y. 2001 ApJ 549, L191

\bibitem[Mihos, Spaans, \& McGaugh(1999)] {mihos99}
Mihos, C., Spaans, M., \& McGaugh, S. 1999 ApJ 515, 89

\bibitem [McGaugh \& Bothun(1994)] {mcgaugh94}
McGaugh, S. \& Bothun, G. 1994 AJ 107, 530

\bibitem [McGaugh(1994)] {mcgaugh94b}
McGaugh, S. 1994 ApJ 426, 13

\bibitem[Nilson(1973)]{nilson73}
Nilson, P. 1973 {\it Uppsala General Catalogue of Galaxies} 
Uppsala Astronomiska Observatoriums Annaler (Uppsala: Astronomiska Observatorium)

\bibitem [O'Neil, et.al(2003)]{oneil02}
O'Neil, K., Bothun, G., van Driel, W., \& Monnier-Ragaigne, D.  2003, preprint

\bibitem [O'Neil \& Schinnerer(2003)]{oneil02b}
O'Neil, K. \& Schinnerer, E. 2003, preprint

\bibitem [O'Neil, Hofner, \& Schinnerer(2000)] {oneil00b}
O'Neil, K., Hofner, P., \& Schinnerer, E. 2000 ApJ 545, L99 

\bibitem [O'Neil, Bothun, \& Schombert(2000)] {oneil00}
O'Neil, K., Bothun, G., \& Schombert, J. 2000 AJ 119, 136

\bibitem [O'Neil, Bothun, \& Cornell(1997a)] {OBC97}
O'Neil, K., Bothun, G., \& Cornell M. 1997a AJ 113, 1212

\bibitem [O'Neil, et.al(1997b)] {oneil97}
O'Neil, K., Bothun, G., Schombert, J., Cornell, M., \& Impey C. 1997b AJ 114, 2448

\bibitem[Oke \& Gunn(1983)]{oke83}
Oke, J. \& Gunn, J. 1983 ApJ 266, 713

\bibitem[Oke(1990)]{oke90}Oke 1990, AJ 99, 1621

\bibitem[Pickering, et.al(1997)]{pickering97}
Pickering, T.E., Impey, C.D., van Gorkom, J.H., \& Bothun, G.D. 1997 AJ 114, 1858

\bibitem[Pildis, Schombert, \& Eder(1997)]{pildis97}
Pildis, R., Schombert, J., \& Eder, J. 1997 ApJ 481, 157

\bibitem[Roennback \& Bergvall(1995)]{roennback95}
Roennback, J. \& Bergvall, N. 1995 A\&A 302, 353

\bibitem[Sanders, et.al(1986)] {sanders86}
Sanders, D., Scoville, N., Young, J., Soifer, B., Schloerb, F., Rice, W.,
\& Danielson, G. 1986 ApJ 305, L45

\bibitem[Schlegel, Finkbeiner, \& Davis(1998)]{schl98}
Schlegel, D., Finkbeiner, D., \& Davis, M 1998 ApJ 500, 525

\bibitem[Schombert, et.al(1992)] {schom92}
Schombert, J., Bothun, G., Schneider, S., \& McGaugh, S. 1992 AJ 103, 1107

\bibitem[Schombert, et.al(1990)] {schom90}
Schombert, J., Bothun, Gregory D., Impey, Chris D., \& Mundy, Lee G. 1990 AJ 100, 1523

\bibitem[Schombert(1998)] {schom98}
Schombert, J. 1998 AJ 116, 1650

\bibitem[Schombert \& Bothun(1988)] {schom88}
Schombert, J. \& Bothun, G. 1988 AJ 95, 1389

\bibitem[Seaton(1979)]{seat79}
Seaton, M. 1979 MNRAS 187, 73

\bibitem[Sprayberry, et.al(1995)]{sprayberry95}
Sprayberry, D., Impey, C.D., Bothun, G.D., \& Irwin, M.J. 1995 AJ 109, 558

\bibitem[Tacconi \& Young(1987)]{tacconi87}
Tacconi, L. \& Young, J. 1987 ApJ 322, 681

\bibitem[van Zee, et.al(1998)]{vanz98}
van Zee, L.,  Salzer, J., Haynes, M., O'Donoghue, A., \& Balonek, T. 1998 AJ 116, 2805

\bibitem[Wilson(1995)] {wilson95}
Wilson, C. D. 1995, ApJ 448, L97

\bibitem[Young, et.al(1995)] {young95}
Young, J. et.al 1995 ApJS 98, 219
\end{thebibliography}
\end{document}